\documentclass[prd,floats,superscriptaddress,showpacs,nofootinbib,twocolumn,preprintnumbers]{revtex4}% , 

\usepackage{amssymb}
\usepackage{amsmath}

\usepackage{sans}

\usepackage{graphicx}
\usepackage{graphics}
\usepackage{dcolumn}
\usepackage{color}
\usepackage{rotate}
\usepackage{fancyhdr} 
\usepackage{hyperref}
\usepackage{indentfirst}

\usepackage{umoline}

\def\be{\begin{eqnarray}}
\def\ee{\end{eqnarray}}
\def\ba{\begin{eqnarray}}
\def\ea{\end{eqnarray}}
\def\no{\nonumber}

\definecolor{darkred}{rgb}{.743,0,0}

\begin{document}
\preprint{DESY 19-036}
\title{Ultra-light dark matter in disk galaxies}

\author{Nitsan Bar}\email{nitsan.bar@weizmann.ac.il}\affiliation{Weizmann Institute of Science,
  Rehovot 7610001, Israel} 

\author{Kfir Blum}\email{kfir.blum@cern.ch}\affiliation{Weizmann Institute of Science,
  Rehovot 7610001, Israel}
\affiliation{Theory department, CERN, CH-1211
  Geneve 23, Switzerland} 
  
\author{Joshua Eby} \email{joshaeby@gmail.com}\affiliation{Weizmann Institute of Science,
  Rehovot 7610001, Israel}
  
\author{Ryosuke Sato}\email{ryosuke.sato@desy.de}\affiliation{
Deutsches Elektronen-Synchrotron (DESY), Notkestra{\ss}e 85,  D-22607 Hamburg, Germany
}

\date{\today}

\begin{abstract}
Analytic arguments and numerical simulations show that bosonic ultra-light dark matter (ULDM) would form cored density
distributions (``solitons'') at the center of galaxies. 
ULDM solitons offer a promising way to exclude or detect ULDM by looking for a distinctive feature in the central region of galactic rotation curves. Baryonic contributions to the gravitational potential pose an obstacle to such analyses, being (i) dynamically important in the inner galaxy and (ii) highly non-spherical in rotation-supported galaxies, resulting in non-spherical solitons. 
We present an algorithm for finding the ground state soliton solution in the presence of stationary non-spherical background baryonic mass distribution. 
We quantify the impact of baryons on the predicted ULDM soliton in the Milky Way and in low surface-brightness galaxies from the SPARC  database.
\end{abstract}
%\pacs{}

\maketitle

%\tableofcontents

%%%%%%%%%%%%%%%%%%%
\section{Introduction}
An ultra-light bosonic field oscillating around a minimum of its potential~\cite{Hu:2000ke,Svrcek:2006yi,Arvanitaki:2009fg,Marsh:2015xka} can play the role of dark matter (DM).  
On cosmologically large scales ultra-light dark matter (ULDM)
behaves similarly to cold weakly-interacting massive particle (WIMP) dark matter, reproducing its success with respect to the cosmic microwave background and large-scale structure. On smaller scales comparable to the de Broglie wavelength, ULDM behaves differently to WIMPs. In particular, at the centre of galactic halos ULDM develops cored density profiles that lead to markedly different predictions than those found for ordinary WIMPs~\cite{Hu:2000ke,Arbey:2001qi,Lesgourgues:2002hk,Chavanis:2011zi,Chavanis:2011zm,Schive:2014dra,Schive:2014hza,Marsh:2015wka,Calabrese:2016hmp,Chen:2016unw,Schwabe:2016rze,Veltmaat:2016rxo,Hui:2016ltb,Gonzales-Morales:2016mkl,Robles:2012uy,Bernal:2017oih,Mocz:2017wlg,Mukaida:2016hwd,Vicens:2018kdk,Bar:2018acw,Eby:2018ufi,Bar-Or:2018pxz,Marsh:2018zyw,Chavanis:2018pkx,Emami:2018rxq,Levkov:2018kau,Broadhurst:2019fsl}. The cored ULDM distributions correspond to quasi-stationary minimum energy solutions of the equations of motion. We will follow common convention and refer to these solutions as ``solitons".

Ref.~\cite{Bar:2018acw} analysed the rotation curves of well-resolved low surface-brightness (LSB) disk galaxies from the SPARC database~\cite{Lelli:2016zqa} and pointed out that these galaxies fail to show the soliton feature predicted by numerical simulations~\cite{Schive:2014dra,Schive:2014hza,Veltmaat:2018dfz}\footnote{Ref.~\cite{Deng:2018jjz} reported independent evidence against soliton cores.}. This led to the bound $m\gtrsim10^{-21}$~eV. A similar constraint\footnote{Ref.~\cite{Marsh:2018zyw} noted that dynamics of a central star cluster in Eridanus-II could potentially probe ULDM up to $m\sim10^{-19}$~eV.} was found in~\cite{Marsh:2018zyw} considering the dwarf spheroidal galaxy Eridanus-II. The matter power spectrum inferred from Ly-$\alpha$
forest analyses yields a comparable bound~\cite{Armengaud:2017nkf,Irsic:2017yje,Zhang:2017chj,Kobayashi:2017jcf,Leong:2018opi}\footnote{See also~\cite{Bozek:2014uqa,Hlozek:2017zzf}. A  bound, $m\gtrsim10^{-23}$~eV, comes from scalar metric perturbations induced by ULDM~\cite{Khmelnitsky:2013lxt} that were searched for in pulsar timing signals~\cite{Porayko:2018sfa}. Heating of the MW disk suggests $m>0.6\times10^{-22}$~eV~\cite{Church:2018sro}. More tentative constraints  include $m>1.5\times10^{-22}$~eV~\cite{Amorisco:2018dcn}, based on preliminary analysis of stellar streams in the Milky Way, and $m>8\times10^{-21}$~\cite{Schneider:2018xba}, assuming that 21cm results by EDGES~\cite{Bowman:2018yin} are confirmed.}. These lower bounds on $m$ are interesting because they probe DM using gravity alone, without requiring any direct interactions with SM fields; because they define how light DM could possibly be; and also because ULDM with $m\sim (10^{-22} - 10^{-21})$~eV was suggested as an explanation for puzzles facing the WIMP paradigm
on small scales~\cite{DelPopolo:2016emo,Hui:2016ltb}. 

In the attempt to constrain (or detect) ULDM with galactic kinematics, an important issue is the modelling of the baryonic contribution to the gravitational potential which can distort the soliton\footnote{See~\cite{2017arXiv171201947C} for a preliminary study of the dynamical impact of stars in ULDM numerical simulations.}. 
Ref.~\cite{Bar:2018acw} analysed the solution in the presence of a spherically-symmetric background potential, in order to estimate the size of the effect. That was found to be significant for the Milky Way (MW), but not significant for the relevant SPARC LSB galaxies. However, both in the MW and in rotation-supported SPARC galaxies, the baryonic mass distribution is non-spherical, following disk-like morphology. In a non-spherical system dynamics in the central region of the galaxy can be affected by the mass distribution at larger radii. It is therefore important to extend the soliton+baryon analysis to non-spherical configurations.  

In this paper we present an algorithm to calculate the soliton solution in the presence of a non-spherical background gravitational potential. The algorithm is simple, fast and accurate and can replace the standard one-dimensional shooting method used for solving the spherically symmetric soliton. 

Our goals in presenting this tool are twofold. 
First, it allows to perform a self-consistent analysis of the velocity profile in disk galaxies. 
Once the baryonic mass distribution is specified (presumably with input from photometry), the soliton contribution to the gravitational potential requires a single free parameter in the fit. This parameter can be chosen to be, e.g., the soliton mass. For example, stellar kinematics in the MW could provide a testing ground for ULDM up to $m\sim10^{-19}$~eV~\cite{Bar:2018acw}. To this end, implementing the soliton in a self-consistent manner would be crucial and we expect that our tool would be useful. 

Second, we revisit the analysis of Ref.~\cite{Bar:2018acw} of baryonic effects in SPARC galaxies. As noted in~\cite{Bar:2018acw}, the soliton--halo relation predicted by DM-only numerical simulations strongly over-predicts the circular velocity in the centres of dozens of galaxies if $m<10^{-21}$~eV. In many cases, the predicted soliton mass in the central $\sim100$~pc of the galaxy exceeds the observationally allowed total mass (baryonic+DM) in that region by factors of order 10. This large mass mismatch led~\cite{Bar:2018acw} to expect that baryonic effects are unlikely to change the constraints. Here, focusing on two sample galaxies, we determine the soliton solution while accounting for the non-spherical baryonic mass distribution. When noting that the gas and stellar  distributions are highly non-spherical, and when naively counting the mass outside of the soliton region, the total baryonic mass in both galaxies is comparable to or larger than the soliton mass. Nevertheless, in both cases our analysis largely confirms the expectations of~\cite{Bar:2018acw}, showing that the baryonic mass external to the soliton region does not significantly affect the solution. 

Some recent work in the literature investigated non-spherical distributions of condensed dark matter \cite{Hayashi:2019ynr,Alexander:2019qsh}. Ref.~\cite{Hayashi:2019ynr} considered a non-spherical parametrisation of ULDM halos and solitons and applied it to a Jeans analysis of dwarf spheroidal galaxies. Differently to our work here, Ref.~\cite{Hayashi:2019ynr} did not base their parametrisation of the ULDM core on a solution of the  equations of motion (EOM). To our view, one of the main points of beauty in the discussion of ULDM in galaxies is that numerical simulations -- with and without stars -- actually do consistently show soliton solutions that satisfy the EOM\footnote{See Secs.~III and~V.A in Ref.~\cite{Bar:2018acw}.}. The tools we present here solve the EOM and find the self-consistent soliton, allowing to refine the analysis of~\cite{Hayashi:2019ynr}. Ref.~\cite{Alexander:2019qsh} looked for disk configurations of self-interacting condensed dark matter. However, while \cite{Alexander:2019qsh} looked for non-spherical configurations, at no point do they solve the EOM. Instead, they restrict the solution to certain disk-like geometries and define and solve a modified 1D system.

The outline of this paper is as follows. 
In Sec.~\ref{s:sol} we recall the ULDM equations of motion that define the soliton solution in the presence of an external (non-dynamical) background gravitational potential. The standard one-dimensional shooting method, that can be used to solve the spherically-symmetric problem, becomes impractical (in general) once the background potential is not spherically-symmetric because it requires a discrete infinity of shooting variables\footnote{If the problem is axisymmetric, for example, then one independent shooting variable is needed for every azimuthal Legendre $l$-mode. %See~\cite{Alexander:2019qsh} for a previous attempt to solve the problem.
}. A simple numerical recipe to solve this problem, assuming an axisymmetric background potential, is detailed in App.~\ref{app:num}. While we do not pursue this here, extending the algorithm to full 3D is straightforward.

In Sec.~\ref{ss:E2MvsK2M} we discuss the soliton--host halo mass relation found in DM-only numerical simulations~\cite{Schive:2014dra,Schive:2014hza,Veltmaat:2018dfz}. Ref.~\cite{Bar:2018acw} showed that this relation is equivalent to the statement, that the specific energy (total energy per unit mass) of the host halo is equal to the specific energy of a self-gravitating soliton. Here we point out that a more physical representation of the soliton--halo relation is obtained by equating the kinetic -- rather than total -- energy per unit mass of the soliton and the halo. This distinction is unimportant for DM-only simulations of massive halos, but becomes relevant once a background potential is introduced.

In Sec.~\ref{s:mw} we take the MW as an illustrative example of a system where the potential due to baryons (mostly stars in this case) cannot be neglected in assessing the soliton properties. The analysis demonstrates the use of our numerical tool, but is not intended to provide constraints on ULDM: that would require a more comprehensive treatment that we postpone to future work. 

In Sec.~\ref{s:sparc} we consider two sample LSB galaxies from the SPARC database. For these galaxies, we reconstruct the circular velocity decomposition presented in the SPARC database using photometric data, reproducing the SPARC analysis. The baryonic mass models (stars+gas) derived in this way are used as input for the numerical non-spherical soliton solution, allowing us to revisit in detail the earlier rough estimates of Ref.~\cite{Bar:2018acw}. We show that the total energy per unit mass, $E/M$, of the soliton is modified by the baryonic potential of these galaxies. However, the bulk of this effect is unphysical: it comes from a non-dynamical shift of the energy due to an external gravitational potential that is mostly constant throughout the relevant region of the galaxy. This is supported by the fact that -- as we show -- the specific kinetic energy, $K/M$, is essentially unaffected both for the soliton and the halo.

In Sec.~\ref{s:sum} we summarise our results.

A number of technical details are postponed to appendices. As mentioned above, App.~\ref{app:num} describes the non-spherical soliton-finding algorithm. App.~\ref{app:bh} specifies the steps required to implement a black hole in the code. In App.~\ref{app:theory} we discuss tests of the algorithm and show evidence that the solutions we find are indeed ground-state solutions. In App.~\ref{app:disk} we recall a convenient formula converting an axisymmetric mass distribution into the gravitational potential induced by it. In Sec.~\ref{app:mnfits} we collect a useful auxiliary parametrisation for galactic discs, that we have found useful in modelling SPARC galaxies. App.~\ref{app:ugc1281gas} explains our reconstruction of the neutral gas distribution in UGC01281. Finally, our results in the main text are presented -- for concreteness -- assuming ULDM particle mass of $m=10^{-22}$~eV; in App.~\ref{app:1e-21} we show relevant results for $m=10^{-21}$~eV.

\section{Solitons in a non-spherical background}\label{s:sol}
We consider a real, massive, free\footnote{Analyses of interacting fields can be found in,
  e.g.~\cite{Chavanis:2011zi,Chavanis:2011zm,RindlerDaller:2012vj,Desjacques:2017fmf}.}
scalar field $\phi$ satisfying the Klein-Gordon equation of motion
and minimally coupled to gravity.   
In the non-relativistic regime it is convenient to decompose $\phi$ as
\be\label{eq:schroedfield}\phi(  x,t)=\frac{1}{\sqrt{2}m}e^{-imt}\psi(  x,t)+c.c.,\ee
with complex field $\psi$ that varies slowly in space and time and
satisfies the Schr\"odinger-Poisson equations (SPE)~\cite{Ruffini:1969qy}
\be\label{eq:SP1} i\partial_t\psi&=&-\frac{1}{2m}\nabla^2\psi+m\left(\Phi+\Phi_b\right)\psi,\\
\label{eq:SP2}\nabla^2\Phi&=&4\pi G|\psi|^2.\ee
In Eq.~(\ref{eq:SP1}) we include an external contribution to the gravitational potential, given by $\Phi_b$. We consider $\Phi_b$ as the effect of a distribution of baryonic mass. Our working assumption is that $\Phi_b$ should be constrained by external information such as photometry and microlensing measurements. 

We look for a quasi-stationary phase-coherent solution for the ULDM, described by the ansatz
\be\label{eq:schroedfield2}\psi(  x,t)&=&\left(\frac{mM_{pl}}{\sqrt{4\pi}}\right)e^{-i\gamma mt}\chi(  x)\ee
where $M_{pl}=1/\sqrt{G}$. 
The parameter $\gamma$ is an eigenvalue of the SPE subject to the bound-state boundary conditions that we describe below. 

We rescale the spatial coordinate, 
\be  x\to m  x,\ee
keeping this convention throughout the rest of the paper. Then in terms of the dimensionless $\chi$ and $  x$ the SPE are
\begin{align}
\nabla^2 \chi &= 2(\Phi + \Phi_b - \gamma )\chi,\label{eq:schrodinger}\\
\nabla^2 \Phi &= \chi^2. \label{eq:poisson}
\end{align}

We assume cylindrical symmetry and parity symmetry ($x_3 = z\to -z$), and define the radial coordinate in the plane $R=\sqrt{x_1^2+x_2^2}$. At $\sqrt{R^2+z^2}\to\infty$ the potentials $\Phi$ and $\Phi_b$ are assumed to decay $\propto1/\sqrt{R^2+z^2}$, implying that $\chi$ decays approximately exponentially $\propto e^{-\sqrt{2|\gamma|(R^2+z^2)}}$. 
A given value of $\chi$ at the origin, specified by 
\be\label{eq:lam}\chi(R=0,z=0)=\lambda^2\ee
with $\lambda$ a real positive number, fixes the minimal energy solution of Eqs.~(\ref{eq:schrodinger}-\ref{eq:poisson}) consistent with the boundary conditions. 

In the case of vanishing $\Phi_b$, solutions of Eqs.~(\ref{eq:schrodinger}-\ref{eq:poisson}) admit a scaling symmetry, the orbit of which can be parametrised by $\lambda$. This scaling symmetry is, in general, broken by $\Phi_b\neq0$. It remains true, however, that varying the value of $\lambda$ in Eq.~(\ref{eq:lam}) generates the continuous family of solutions of Eqs.~(\ref{eq:schrodinger}-\ref{eq:poisson}). Thus, $\lambda$ remains a useful tool to parameterise the mass, energy and any other property of the solution. For reference, the self-gravitating soliton (found for $\Phi_b=0$) satisfies $M\approx2.06\lambda\frac{M_{pl}^2}{m}\approx2.8\times10^{12}\lambda\left(\frac{m}{10^{-22}~\rm eV}\right)^{-1}$~M$_{\odot}$ and $E/M\approx-0.23\lambda^2\approx-0.054\left(\frac{M}{M_{pl}^2/m}\right)^2$. When baryons induce $\Phi_b\neq0$ these relations are modified in a way that we will discuss below.\\

We have developed a numerical relaxation method to find the ground state soliton solution for any axisymmetric background potential satisfying the boundary conditions described below Eq.~(\ref{eq:poisson}). The algorithm is described in App.~\ref{app:num}, and is one of the main results of this paper. We discuss some theoretical aspects of the solutions in App.~\ref{app:theory}: first, the evidence for (but difficulty to rigorously prove) that the solution is indeed the ground state, and second the issue of stability against small perturbations.

In the next subsection we clarify some issues related to the soliton--host halo relation found in DM-only numerical simulations. 
Then, in the following sections we illustrate the use of the numerical tool of App.~\ref{app:num} by analysing the baryonic effects on the predicted ULDM soliton in the Milky Way and in two disk galaxies from the SPARC  database.

\subsection{Soliton -- halo relation: $E/M$ vs. $K/M$}\label{ss:E2MvsK2M}

We can compute the soliton mass $M$ and energy $E$ (recall that $ x$ is measured in units of $1/m$),
\be
\label{eq:M}M &\approx&10^{11}{~\rm M_\odot}\left(\frac{m}{10^{-22}~\rm eV}\right)^{-1}\int d^3x\,\chi^2,
\ee
\be\label{eq:E}
E &\approx& 10^{11}{~\rm M_\odot}\left(\frac{m}{10^{-22}~\rm eV}\right)^{-1}\,\times\\
&&\int d^3x\left(\frac{1}{2}\left(\nabla \chi\right)^2+\left(\dfrac{\Phi}{2}+\Phi_b\right)\chi^2\right).\no
\ee
It is useful to separate the total energy into kinetic energy + potential energy,
\be E&=&K+P,\ee
where $K$ comes from the gradient term and $P$ comes from the $\Phi/2+\Phi_b$ term in Eq.~(\ref{eq:E}). 
For a self-gravitating system in virial equilibrium, $P=-2K$ and $E=-K$. This applies to the self-gravitating soliton obtained for $\Phi_b=0$. When we turn on a background potential the soliton ceases to be self-gravitating, so that $E\neq-K$ for $\Phi_b\neq0$. 

Ref.~\cite{Bar:2018acw} showed that the empirical soliton-host halo relation found in the DM-only numerical simulations of Ref.~\cite{Schive:2014hza} is equivalent to the statement
\be\label{eq:solhal}
\dfrac{E}{M}\Big|_{\rm soliton}&=&\dfrac{E}{M}\Big|_{\rm halo}.
\ee
Note that on the LHS of Eq.~(\ref{eq:solhal}), $\dfrac{E}{M}\Big|_{\rm soliton}$ is defined for the self-gravitating soliton {\it without} including the gravitational potential induced by the large-scale halo. The halo gravitational potential $\Phi_h$ is approximately constant in the halo inner region where the soliton occurs and can be estimated as $\Phi_h\sim10\dfrac{E}{M}\Big|_{\rm halo}$, up to $\mathcal{O}(1)$ corrections depending on the detailed shape of the halo~\cite{Bar:2018acw}. If we were to include the correction to the soliton energy due to this constant background potential, it would change: $\dfrac{E}{M}\Big|_{\rm soliton}\to\dfrac{E}{M}\Big|_{\rm soliton}+\Phi_h$. This discussion suggests that the soliton--host halo relation is better expressed using kinetic energy, rather than total energy:
\be\label{eq:solhalK}
\frac{K}{M}\Big|_{\rm soliton}&=&\frac{K}{M}\Big|_{\rm halo}.
\ee
Because $\Phi_h$ is approximately constant over the region where the soliton is supported, the soliton shape is not distorted and its kinetic energy is not modified from its value for the self-gravitating solution. This means that for massive halos in DM-only simulations, Eq.~(\ref{eq:solhalK}) and Eq.~(\ref{eq:solhal}) are indistinguishable.

Eq.~(\ref{eq:solhalK}) and Eq.~(\ref{eq:solhal}) become distinguishable when we turn on $\Phi_b\neq0$, with a nontrivial spatial profile such that $\Phi_b$ is not constant throughout the large-scale halo. 

\section{Application: The Milky Way}\label{s:mw}

We now consider soliton solutions in the background of a gravitational potential $\Phi_b$, chosen to roughly mimic the inner region of the MW. Our goal is to illustrate the approximate size of the baryonic effects on the soliton, and not to characterise these effects in full; a detailed, accurate and precise modelling of the inner MW stellar and gas mass distributions is challenging and is postponed to future work. For concreteness, throughout this section we set $m=10^{-22}$~eV. 

The dominant contributions to the stellar mass profile of the MW inner few hundred pc were described in the photometric analysis of Launhardt et al~\cite{Launhardt:2002tx} as a spherical nuclear stellar cluster (NSC) and a nuclear stellar disk (NSD), composing together the nuclear bulge (NB).

In addition to the stellar components, dynamics in the central $\sim1$~pc is dominated by a super-massive black hole (SMBH) with mass $M_{BH}\approx4\times10^6$~$M_\odot$. Here we omit the SMBH contribution, which was studied in~\cite{Bar:2018acw} and shown to have negligible impact on the soliton for $m\lesssim10^{-20}$~eV. We note that the numerical code in App.~\ref{app:num} is capable of handling the SMBH contribution via the procedure described in App.~\ref{app:bh}. A gas torus at scale radius of $\sim100$~pc contributes $\sim2\times10^7$~M$_\odot$. For simplicity, the gas is also neglected here in comparison to the stellar components. 

The NSC density profile was modelled as
\be \rho_{NSC}(r)&=&\frac{\bar\rho_{NSC}}{1+\left(\frac{r}{0.22}\right)^{n_{NSC}}}\theta\left(200-r\right),\ee
where $r=\sqrt{R^2+z^2}$ is stated in pc. $\bar\rho_{NSC}=3.3\times10^6$~$M_\odot$/pc$^3$ for $ r<r_0 $ and $\bar\rho_{NSC}=9.0\times10^7$~$M_\odot$/pc$^3$ for $ r\ge r_0 $, with $ r_0=6 $~pc. The index $n_{NSC}=2$ for $r<r_0$ and $n_{NSC}=3$ for $r\geq r_0$ (keeping the profile continuous at $r_0$). With these parameters we have\footnote{This NSC mass is larger than that quoted in~\cite{Launhardt:2002tx} by a factor of $\sim1.8$. We are not sure of the reason for this mismatch, but it does not have an important effect on our results.} $M_{NSC}\simeq 5.3\times10^7$~M$_\odot$.

We parametrise the NSD stellar mass density as follows,  
\be\rho_{NSD}\left(R,z\right)&=&\frac{\bar\rho_{NSD}}{1+\left(\frac{R}{250}\right)^{14}}\left(1-\tanh^4\left(\frac{R}{140}\right)\right)e^{-\frac{|z|}{15}},\no\\&&\label{eq:nsd}
\ee
where $\bar\rho_{NSD}=330$~$M_\odot$/pc$^3$ and where $z$ and $R$ are stated in pc. This parametrisation approximately reproduces the NIR stellar volume emissivity model derived in~\cite{Launhardt:2002tx} and yields an NSD mass $M_{NSD}\simeq10^9$~M$_\odot$, consistent within the uncertainty with the value of $\left(1.4\pm0.6\right)\times10^9$~M$_\odot$ quoted by~\cite{Launhardt:2002tx}.

A kinematic detection supporting the disk-like morphology of the NSD was given in~\cite{2041-8205-812-2-L21}, and the mass and approximate scale estimates are consistent with the dynamical modelling of~\cite{Portail:2016vei} and with microlensing analyses~\cite{2017ApJ...843L...5W} that probe the outer boundary of the NSD region.

In what follows we define $\Upsilon_L \equiv \Upsilon/\Upsilon_{\rm Launhardt}$ as the mass-to-light ratio of the stellar distribution compared to the one used in Ref.~\cite{Launhardt:2002tx}. We vary $\Upsilon_L$ to explore the consequences of different total stellar mass in the NB region.

\begin{figure}[htbp!]
	\centering
	\includegraphics[width=0.45\textwidth]{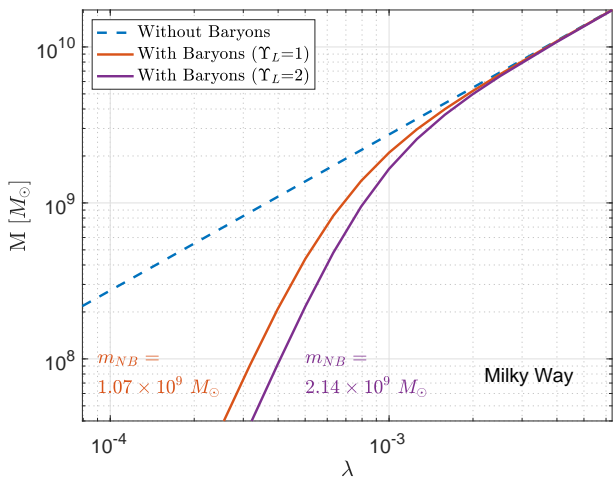}
	\caption{Soliton M-$\lambda$ relation in the stellar-induced background gravitational potential of the inner MW. For a halo mass $M_h=10^{12}$~M$_\odot$, the soliton host-halo relation found in DM-only numerical simulations predicts $\lambda=4.9\times 10^{-4}$. The ULDM particle mass is $m=10^{-22}$~eV. $ \Upsilon_L $ is defined in the text.}\label{fig:MlamMW}
\end{figure}

In Fig.~\ref{fig:MlamMW} we plot the soliton mass vs. $\lambda$, which allows us to access different solutions. For $\lambda\gtrsim10^{-3}$ we retrieve the self-gravitating soliton result, shown by the dashed line. For smaller $\lambda$ we find $M\propto \lambda^4$~\cite{Bar:2018acw}\footnote{\label{fn:1}This can be understood as follows. For small $\lambda$ the external potential dominates and the SPE reduce to $ \nabla^2\chi\approx 2(\Phi_b-\gamma)\chi $. Since this equation is homogeneous and linear in $\chi$, the normalisation at $ x=0$ is a multiplicative factor and $M\propto \int d^3x \,\chi^2\propto \lambda^4$.}. Fig.~\ref{fig:MlamMW} can be compared to Fig.~16 in Ref.~\cite{Bar:2018acw} which considered a spherically-averaged approximation to the same stellar mass model. It shows an $\mathcal{O}(1)$ difference in the $M$ vs. $\lambda$ relation in the phenomenologically interesting range $\lambda\sim10^{-4}-10^{-3}$. 

\begin{figure}[htbp!]
	\centering
		\includegraphics[width=0.45\textwidth]{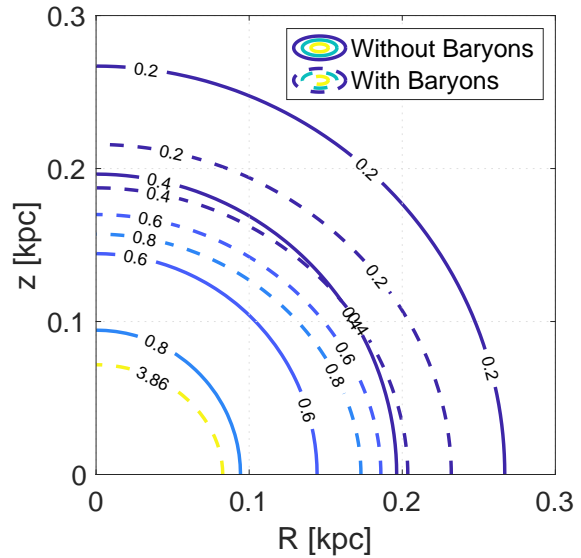}	\caption{Mass density contours of a soliton in the inner MW. The density is normalised to a reference value of 23.6~M$_{\odot}$/pc$^3$. We set $m=10^{-22} $~eV in the plot. The soliton mass is fixed at $M\approx 1.35\times10^{9}~{\rm M}_{\odot}$. Solid lines show the result for the self-gravitating soliton and dashed lines show the result when $\Phi_b$ is included in the SPE.}\label{fig:MWsolitons222}
\end{figure}

\begin{figure}[htbp!]
	\centering
	\includegraphics[width=0.45\textwidth]{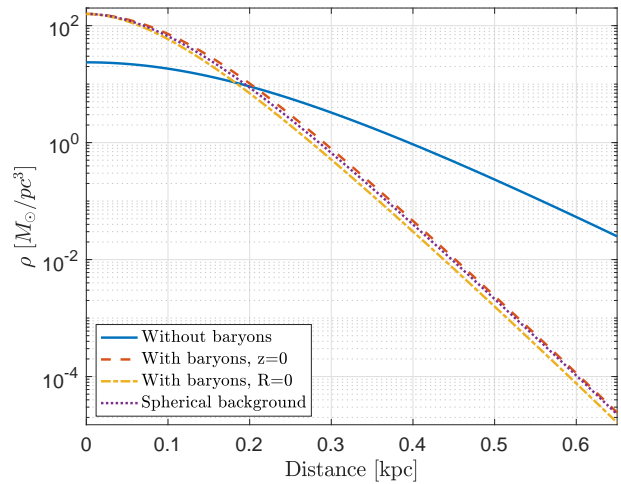}	\caption{ Mass density of solitons corresponding to the inner MW. We set $m=10^{-22} $~eV in the plot. The soliton mass is fixed at $M\approx 1.35\times10^{9}~{\rm M}_{\odot}$. The solid line shows the result for the self-gravitating soliton and the dashed lines show the results when $\Phi_b$ is included in the SPE. The dotted line shows the result when the NSD is replaced by a spherical rearrangement of the same mass.}\label{fig:MWsolitonsDensComp}
\end{figure}

In Fig.~\ref{fig:MWsolitons222} we study the deformation in the soliton shape caused by the stellar mass distribution, at fixed soliton mass $M\approx1.35\times10^9$~M$_\odot$ predicted by DM-only numerical simulations for a halo mass $M_h=10^{12}$~M$_\odot$. The contour lines show the soliton mass density normalised to a reference value of 23.6~M$_\odot$/pc$^3$. Solid lines show the result for the self-gravitating soliton and dashed lines show the result obtained when $\Phi_b$ is included in the SPE. In Fig.~\ref{fig:MWsolitonsDensComp} we plot the density profile of the deformed soliton from Fig.~\ref{fig:MWsolitons222} on the plane of the disk ($ z=0 $, dashed line) and along the $ z $-axis ($ R=0 $, dot-dashed). The solid line shows the density profile of the self-gravitating soliton. The dotted line shows the density profile of the soliton when the baryonic potential is replaced by a radially-averaged version of the potential\footnote{Specifically, we define the spherical rearrangement via $ M(r) = \int_0^r \rho(\mathbf{x})d^3x  $, $ \phi(r) = -\int_r^\infty dr^{\prime} GM(r^{\prime})/(r^{\prime})^2  $.}.

\begin{figure}[htbp!]
	\centering
	\includegraphics[width=0.45\textwidth]{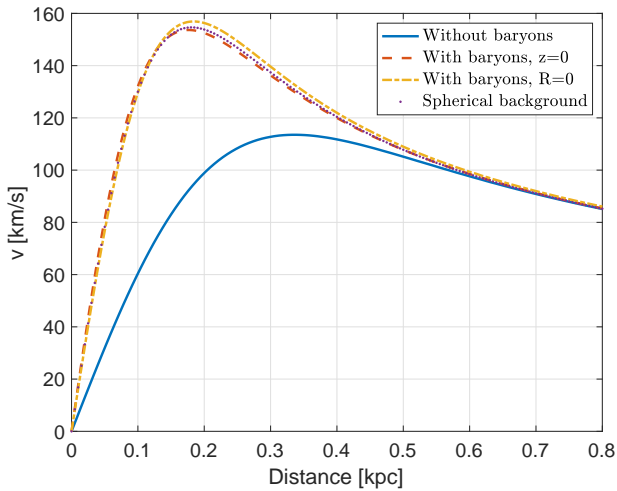}
	\caption{Effective circular velocity induced by a MW soliton. We set $m=10^{-22} $~eV in the plot and fix the soliton mass at $M\approx 1.35\times10^{9}~{\rm M}_{\odot}$, predicted by DM-only numerical simulations for a halo mass $M_h=10^{12}$~M$_\odot$. The solid line shows the result for the self-gravitating soliton and the dashed lines show the result when $\Phi_b$ is included in the SPE. The dotted line shows the result when the NSD is replaced by a spherical rearrangement of the same mass.}\label{fig:MilkyWayVelocity22}
\end{figure}

It is instructive to consider the observable (in principle) soliton-induced effective circular velocity,
\be
v_{\rm eff}(  x)&=&\sqrt{  x\cdot  \nabla\Phi}.\ee
In Fig.~\ref{fig:MilkyWayVelocity22} we plot $v_{\rm eff}$, analogously to Fig.~\ref{fig:MWsolitonsDensComp}. The dashed line is $v_{\rm eff}$ on the plane of the disk. The dot-dashed line is $v_{\rm eff}$ transverse to the disk on the $z$-axis. For comparison, we also plot $v_{\rm eff}$ computed for a self-gravitating soliton with the same mass (solid blue). The main effect of the background stellar potential is to contract the soliton-induced peak velocity deeper into the inner halo, enhancing the peak velocity; this is an $\mathcal{O}(1)$ effect that cannot be ignored in realistic modelling of kinematic data. The deviation from radial symmetry is, however, small: a simplified treatment taking as input a radially-averaged baryonic mass distribution could suffice for practical purposes. For comparison, the result of such a procedure is plotted in the dotted line in Fig.~\ref{fig:MilkyWayVelocity22}.

\begin{figure}[htbp!]
	\centering
	\includegraphics[width=0.495\textwidth]{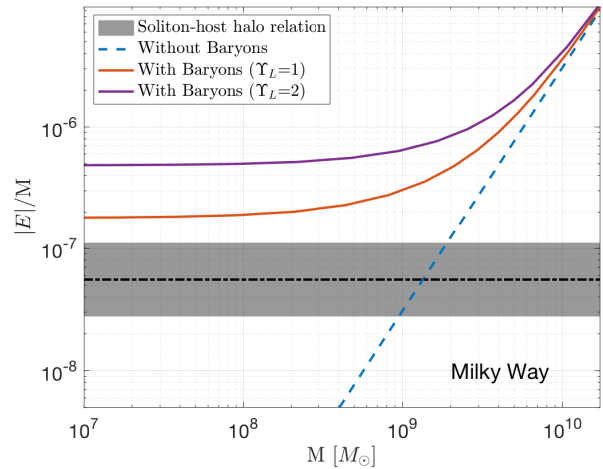}	
	\includegraphics[width=0.495\textwidth]{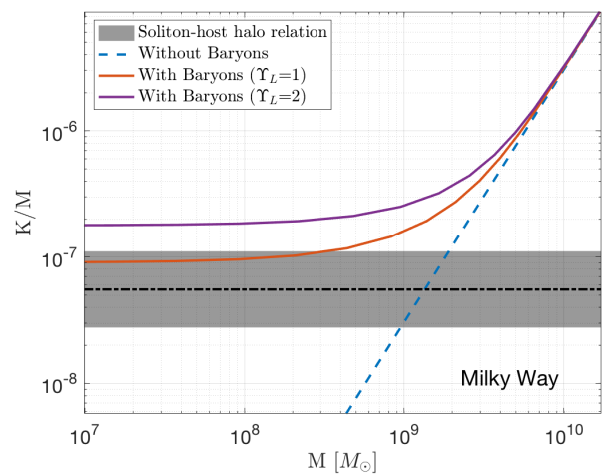}
	\caption{Specific energy $|E|/M$ (top) and specific kinetic energy $K/M$ (bottom) for a soliton in the MW. For a halo mass $M_h=10^{12}$~M$_\odot$, the soliton host-halo relation found in DM-only numerical simulations predicts $|E|/M=K/M\approx5.5\times10^{-8}$~\cite{Bar:2018acw}, shown by the black dot-dashed line with a shaded band denoting a factor of two spread (see text for more details).}\label{fig:MKMW}
\end{figure}

In the top (bottom) panel of Fig.~\ref{fig:MKMW} we plot the total energy (kinetic energy) per unit mass as a function of soliton mass $M$. For $M \gtrsim10^{10}$~M$_\odot$ the self-gravitating soliton result is retrieved. For small $M$ we find that $E/M$ and $K/M$ approach constant values. The reason for this scaling follows along the same lines of footnote~\ref{fn:1} which shows that at small $\lambda$, when the background potential dominates, $M$, $E$ and $K$ all scale as $\propto\lambda^4$ leading to constant $E/M$ and $K/M$.

For a halo mass $M_h=10^{12}$~M$_\odot$, the soliton--host halo relation found in DM-only numerical simulations of~\cite{Schive:2014dra,Schive:2014hza} (summarised by Eq.~(\ref{eq:solhalK})) predicts $K/M\approx5.5\times10^{-8}$~\cite{Bar:2018acw}, shown by the black dot-dashed line. The shaded band denotes a factor of two spread around this prediction, motivated by the  halo-to-halo spread seen in the simulations. 

Fig.~\ref{fig:MKMW} shows that because of the stellar-induced background potential, $K/M$ for an actual soliton solution in this background is significantly deformed. 
This means that baryonic effects are likely to significantly modify the soliton properties, and the soliton--halo expectation from DM-only numerical simulations cannot be taken at face value. A consistent way to constrain (or possibly detect) an ULDM soliton in the MW, would be by a combined analysis of kinematical modelling and photometry, where the stellar potential constrained by photometry is used to self-consistently calculate the soliton shape and where the soliton mass is taken as a free parameter. 

Ref.~\cite{DeMartino:2018zkx} argued for dynamical evidence in favour of an ULDM soliton in the MW, with $m\approx10^{-22}$~eV and $M\approx10^9$~M$_\odot$ in tantalising agreement with the expectations of DM-only numerical simulations. The dynamical evidence for a dense central mass component is consistent with earlier studies~\cite{Launhardt:2002tx,2041-8205-812-2-L21,Portail:2016vei,Bar:2018acw}. Unfortunately, as we reviewed here and in~\cite{Bar:2018acw} (see Sec.~V.B there), there is room for and photometric evidence of about $10^9$~M$_\odot$ in stars within the $\sim200$~pc would-be soliton region~\cite{Launhardt:2002tx}. Thus, the central mass component could well be due to ordinary baryonic matter. 
Other systems, such as well-resolved LSB galaxies, offer much cleaner laboratories in which to look for ULDM solitons. We turn to such systems in the next section.

\section{Application: low surface-brightness SPARC galaxies}\label{s:sparc}

Our second discussion of non-spherical solitons involves two low surface-brightness (LSB) disk galaxies from the SPARC  database~\cite{Lelli:2016zqa}: UGC01281 and F571-8. We choose these galaxies as representative examples of a larger sample including dozens of well-resolved LSB galaxies. For concreteness, throughout this section we set $m=10^{-22}$~eV. Results for $m=10^{-21}$~eV are collected in App.~\ref{app:1e-21}.

The baryonic mass contributions in SPARC galaxies is divided into a spherical bulge component and axisymmetric disk and gas components. The stellar mass distribution is calibrated to match surface brightness data from Spitzer. The computation of the gravitational potential due to the disk is detailed in App.~\ref{app:mnfits}. We focus here on galaxies that are consistent with negligible bulge. 

The gas mass distribution for UGC01281 (not relevant for F571-8) is calibrated to approximately match the HI surface brightness data reported in~\cite{Kamphuis:2011qg}, normalising to the total gas mass reported in~\cite{deBlok:2002vgq}. We provide details on the gas fitting procedure in App.~\ref{app:ugc1281gas}. 

In our computation we fix the total gas mass to match the total mass inferred from the photometry and vary the stellar mass-to-light ratio of the disk from $\Upsilon_d=0$ up to larger values that saturate the observed  kinematic velocity~\cite{Starkman2018}.

\begin{figure}[htbp!]
	\centering
	\includegraphics[width=0.495\textwidth]{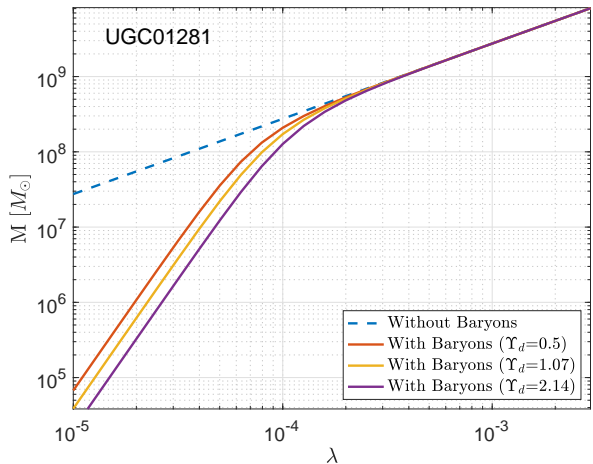}
	\caption{Soliton M-$\lambda$ relation in the baryonic-induced background gravitational potential of UGC01281. The soliton host-halo relation found in DM-only numerical simulations predicts $\lambda=2.2\times 10^{-4}$. The ULDM particle mass is $m=10^{-22}$~eV.}\label{fig:MvsLamugc1281}
\end{figure}

\begin{figure}[htbp!]
\centering
\includegraphics[width=0.495\textwidth]{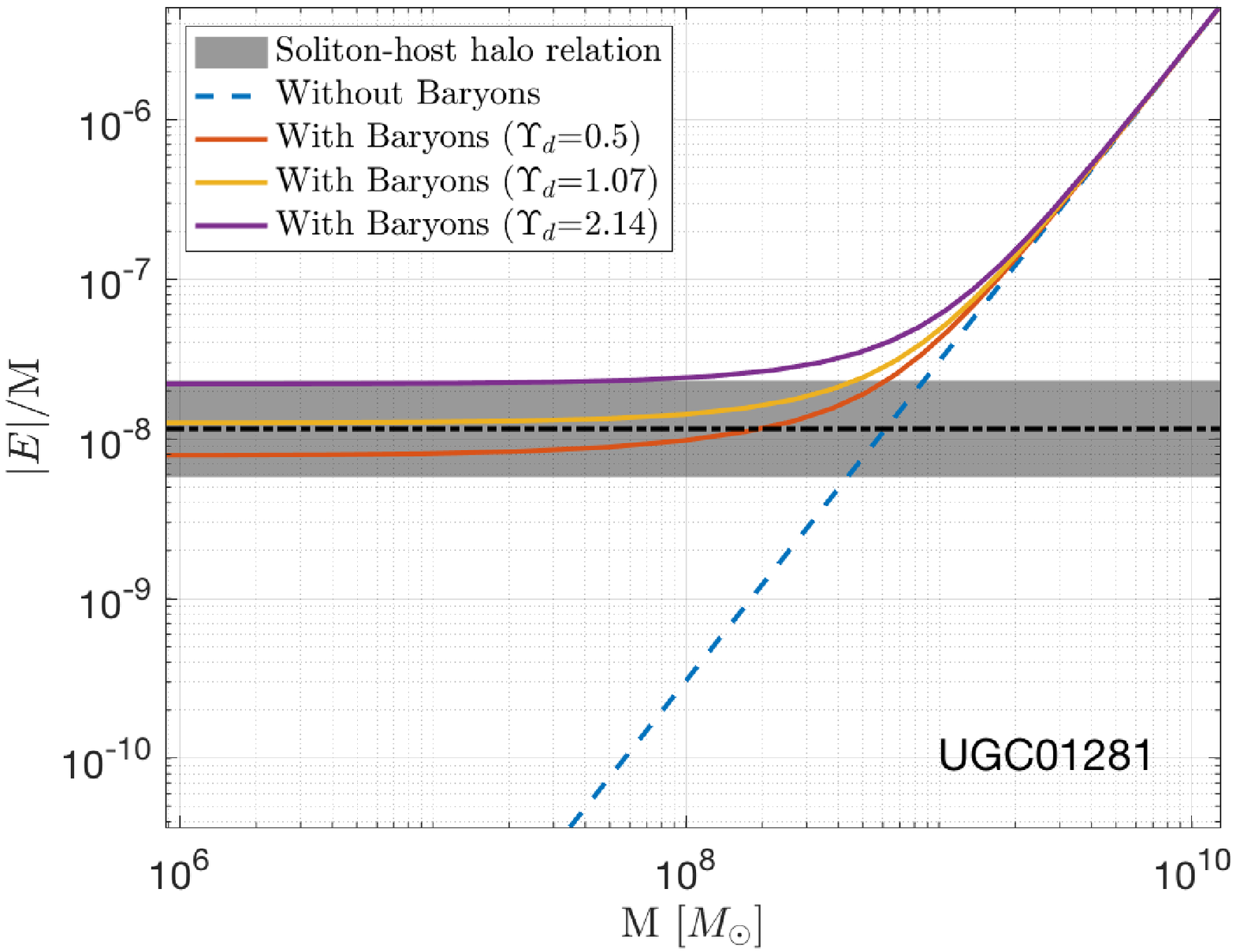}
\includegraphics[width=0.495\textwidth]{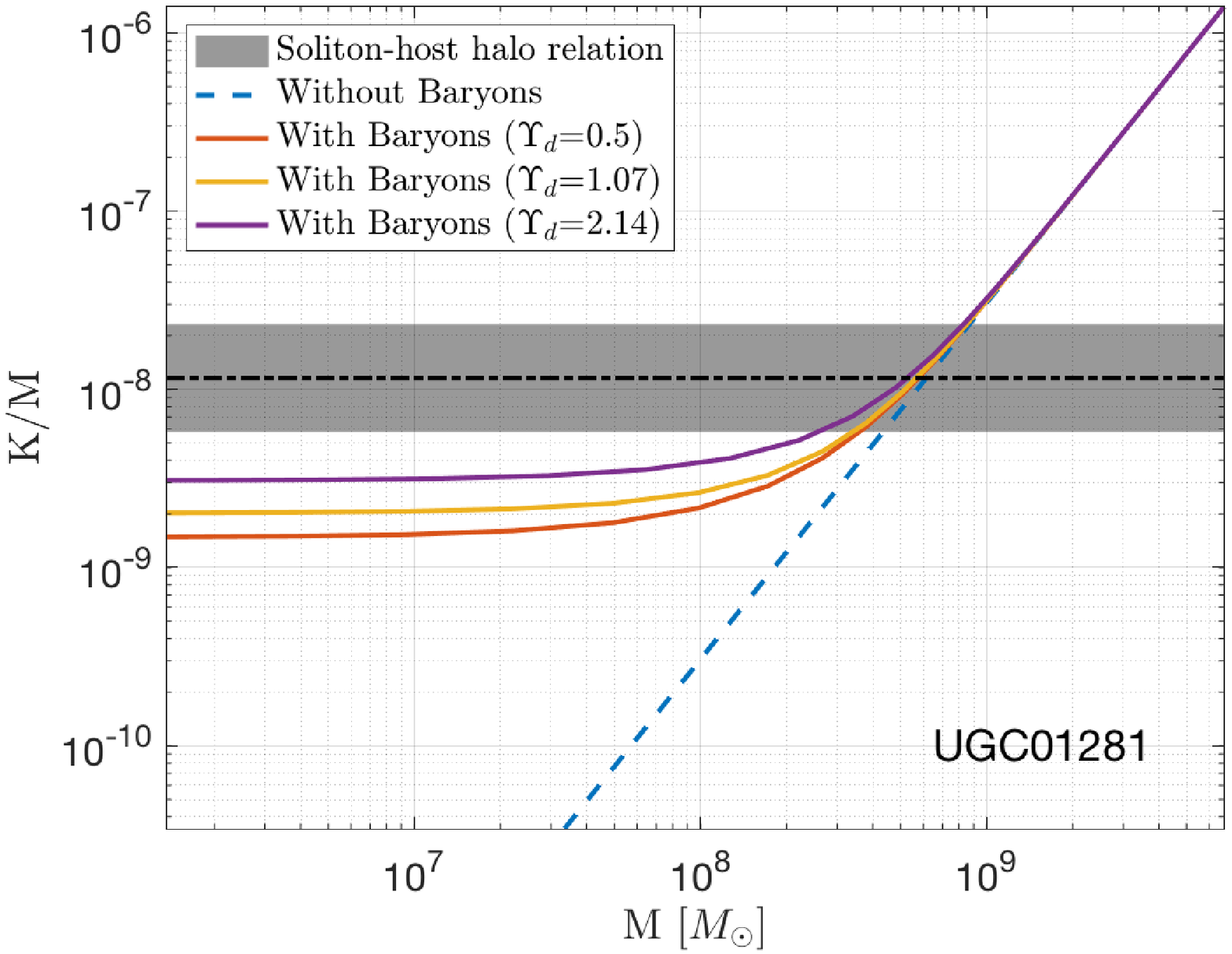}
\caption{Specific energy $|E|/M$ (top) and specific kinetic energy $K/M$ (bottom) for a soliton in UGC01281. The soliton host-halo relation found in DM-only numerical simulations predicts $|E|/M=K/M\approx10^{-8}$~\cite{Bar:2018acw}, shown by the black dashed line with a shaded band denoting a factor of two  spread.}\label{fig:EoverMugc1281}
\end{figure}

In Fig.~\ref{fig:MvsLamugc1281} we plot the $M-\lambda$ relation for a soliton in UGC01281. In the top (bottom) panel of Fig.~\ref{fig:EoverMugc1281} we plot the total energy (kinetic energy) per unit mass vs. $M$. 
The dashed black line denotes the soliton--halo prediction of DM-only numerical simulations. The shaded band shows a factor of two spread around this prediction.

Inspecting Fig.~\ref{fig:EoverMugc1281} we see that in the neighbourhood of $E/M$ values that conform to the DM-only simulation prediction, the actual $E/M$ for a soliton in UGC01281 is significantly shifted compared to the self-gravitating solution. However, the effect on $K/M$ is much less pronounced: the soliton shape is essentially unaffected. 

We can also estimate the baryonic effect on the dynamics of the large-scale halo. To do this, we can compare the observed kinematic velocity at large distances ($r\sim5$~kpc in this example) with the contribution to the velocity that can be attributed to the baryons. The velocity decomposition is shown in the top panel of Fig.~\ref{fig:rcsSPARC} (discussed in more detail at the end of this section). We find $v^2_{\rm baryons}/v^2_{\rm obs}\sim0.26$ ($\sim0.39$), when adopting $\Upsilon_d=1.07$ ($\Upsilon_d=2.14$). This means that the baryonic potential distorts the ULDM large-scale halo $K/M$ by no more than $40\%$.

\begin{figure}[htbp!]
	\centering
	\includegraphics[width=0.495\textwidth]{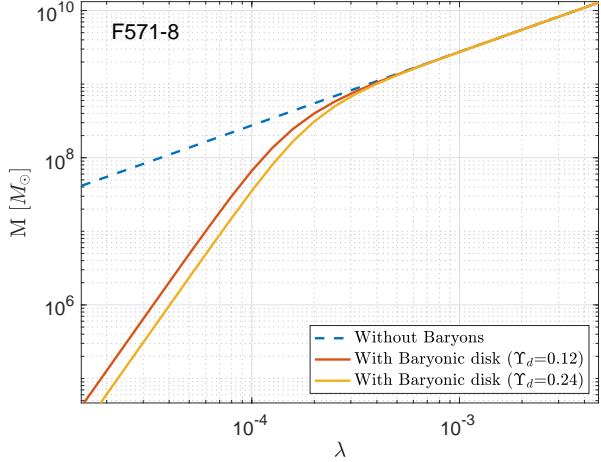}
	\caption{Same as Fig.~\ref{fig:MvsLamugc1281}, but done for F571-8.}\label{fig:MvsLamf571}
\end{figure}

\begin{figure}[htbp!]
\centering
\includegraphics[width=0.495\textwidth]{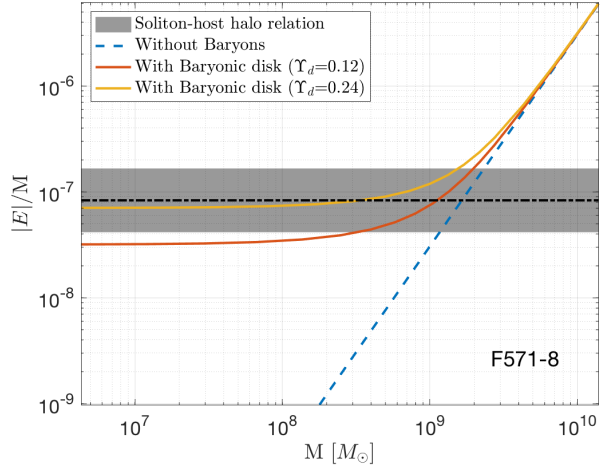}
\includegraphics[width=0.495\textwidth]{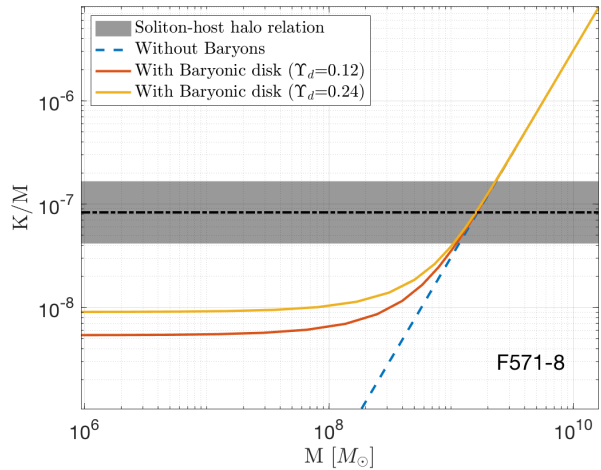}
\caption{Same as Fig.~\ref{fig:EoverMugc1281}, but done for F571-8.}\label{fig:EoverMf571}
\end{figure}

\begin{figure}[htbp!]
	\centering
	\includegraphics[width=0.495\textwidth]{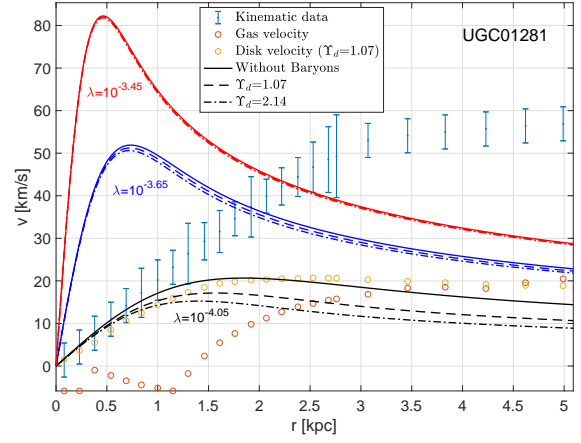}
	\includegraphics[width=0.495\textwidth]{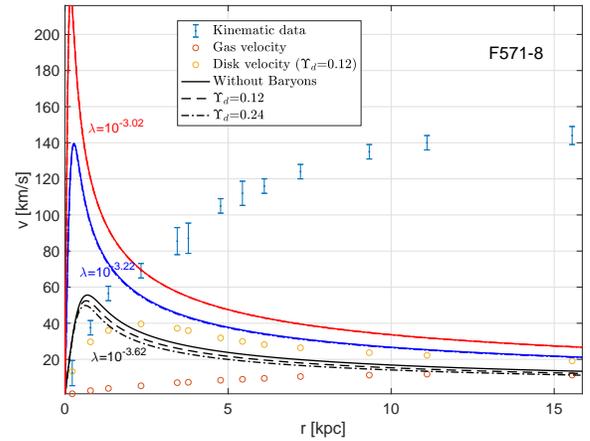}
	\caption{\textbf{Top:} Rotation curve SPARC data of UGC01281, overlaid with soliton solutions assuming background baryonic-induced potential parametrised by stellar disk mass-to-light ratio $\Upsilon_d$ augmented by neutral gas mass distribution consistent with the observed HI brightness measurements. The highest $\Upsilon_d$ is chosen to saturate the error budget of the innermost kinematic velocity data points. The central value, $\lambda=10^{-3.65}$ (blue) is based on the DM-only numerical simulation prediction. 
	\textbf{Bottom:} Same for F571-8 (for this galaxy, the gas contribution is neglected in the soliton computation). The DM-only simulations predict $\lambda=10^{-3.22}$.} \label{fig:rcsSPARC}
\end{figure}

The next galaxy we consider is F571-8.  
Soliton properties for this galaxy are presented in Figs.~\ref{fig:MvsLamf571} and~\ref{fig:EoverMf571}. 
Here, for simplicity, we ignore the (negligible) gas contribution in computing the soliton. 
Again, $K/M$ for a soliton in F571-8 is unaffected by baryons in the parameter region expected from DM-only simulations. The case of F571-8 is even clearer than UGC01281 because the baryonic effect on the dynamics of the large-scale halo, as seen by inspecting the rotation curve decomposition (bottom panel of Fig.~\ref{fig:rcsSPARC}), is not larger than $\sim5\%$.
In the top (bottom) panel of Fig.~\ref{fig:rcsSPARC} we show the rotation curve decomposition of UGC01281 and F571-8, as found in the SPARC  database. The contribution due to soliton solutions with different values of $\lambda$ (indicated in the plot) are overlaid in red, blue and black.  
The solitons are computed assuming different values of the disk stellar mass-to-light ratio $\Upsilon_d$, listed in the caption. 
The central value of $\lambda$ (in blue) corresponds to the prediction of DM-only numerical simulations. For these predicted solitons the baryonic potential makes a negligible impact on the soliton shape, regardless of the value of $\Upsilon_d$ in both galaxies.

We conclude that if Eq.~(\ref{eq:solhalK}) correctly captures the soliton--halo relation of the simulations, then UGC01281 and F571-8 are clean systems in which to constrain the ULDM model, in the sense that the baryonic contribution to the gravitational potential is not important both for the large-scale halo and for the central soliton. These conclusions stay unchanged when we consider more massive ULDM with $m=10^{-21}$~eV (see App.~\ref{app:1e-21}). Dozens of other comparably clean systems exist in the SPARC database. The constraints derived in Ref.~\cite{Bar:2018acw} should therefore apply and ULDM with $m<10^{-21}$~eV is in tension with the data.

%%%%%%%%%%%%%%%%%
\section{Summary}\label{s:sum}

An ultra-light bosonic field oscillating around a minimum of its potential can play the role of dark matter (DM). On scales of order the effective de Broglie wavelength, wave mechanics dictates the dynamics of this ultra-light dark matter (ULDM) opening potential avenues to constrain (or detect) ULDM in various astrophysical and cosmological systems. 

Stellar and gas kinematics of rotation-supported low surface-brightness (LSB) galaxies were used in Ref.~\cite{Bar:2018acw} to derive the constraint $m\gtrsim10^{-21}$~eV. This constraint relies on the validity of a soliton--host halo relation, found in DM-only numerical simulations. It is important to assess to what extent baryons could affect these results. For a non-spherical baryonic distribution, a new numerical tool was required in order to calculate the properties (shape, mass, energetics) of the non-spherical soliton obtained in the presence of the baryonic-induced background gravitational potential. In this paper we provided a simple algorithm (see Sec.~II and App.~A) that achieves this goal.

To illustrate the potential use of the non-spherical soliton solver, we estimated the impact of a Milky Way (MW) nuclear stellar disk (NSD) on an ULDM soliton. Adopting a plausible parameterisation of the stellar distribution, motivated by photometric measurements, we find that the NSD would distort the shape and energetics of an $m=10^{-22}$~eV ULDM soliton at the $\mathcal{O}(1)$ level. Thus, an attempt to constrain ULDM in the MW should self-consistently account for the gravitational effect of stars. While we did not enter such an analysis, the numerical tool we provided is an important step in this direction. Having said that, we note that while the soliton can be compressed by an internal clump of stars it is not easily deformed into non-spherical shape. In the MW example, the highly non-spherical nuclear stellar disk (NSD) leads to a soliton that is significantly contracted but remains spherical to a good approximation. As a result, a spherical rearrangement of the stellar mass distribution (namely, replacing the disk-like baryonic distribution by a radially-averaged profile) would most likely be sufficient to calculate the soliton in a kinematical analysis. 

Next, we revisited the SPARC galaxy analysis of~\cite{Bar:2018acw}. Using two LSB galaxies as a concrete example, we modelled the baryonic potential consistent with photometric data and bracketed the possible impact on the shape and energetics of the predicted soliton. Our results reinforce the conclusions of~\cite{Bar:2018acw}, implying that baryons are not expected to change the constraints derived on ULDM based on rotationally-supported LSB SPARC galaxies.

\acknowledgments
We thank Ben Bar-Or and Scott Tremaine for discussions and Stacy McGaugh for clarifications regarding the SPARC  database. KB is incumbent of the Dewey David Stone
and Harry Levine career development chair. The work of KB and NB was
supported by grant 1937/12 from the I-CORE program of the Planning and
Budgeting Committee and the Israel Science Foundation and by grant
1507/16 from the Israel Science Foundation. The work of JE was supported by the Zuckerman STEM Leadership Program.

\begin{appendix}

\section{Numerical algorithm for solitons in a non-spherical background potential}\label{app:num}
In what follows we describe a numerical method to find the ground state solution of Eqs.~(\ref{eq:schrodinger}-\ref{eq:poisson}). This tool is one of the main results of our work: it is intended to be simple, fast, and robust enough to allow it to be used in detailed analyses of galactic kinematics with ULDM, in cases -- such as the Milky Way galaxy --  where the baryonic contribution to the gravitational potential in the soliton region cannot be neglected. 

We assume that the baryonic-induced gravitational potential is a direct input to the code. Often, an input in terms of the stellar and gas mass density could be more natural. Converting an axisymmetric mass distribution into its corresponding gravitational potential is a straightforward exercise that we recall in App.~\ref{app:disk}.

We use an $N\times N$ discretised lattice with physical size $L \times L$ in the $R-z$ plane.
The lattice spacing is $\delta = L/(N-1)$.
The physical coordinate of each point (recall that distance is measured in units of $1/m$) is
\be
(R_i,z_j) &=& \left( \frac{i-1}{N-1}L,~ \frac{j-1}{N-1}L \right).
\ee
The Laplacian in cylindrical coordinates is
\be
\nabla^2 \Phi 
&=&
\left(\frac{\partial^2}{\partial R^2} + \frac{1}{R} \frac{\partial}{\partial R}\right)\Phi
+
\frac{\partial^2}{\partial z^2}\Phi.
\ee
We discretise it:
\begin{widetext}
\be
\left[\left(\frac{\partial^2}{\partial R^2} + \frac{1}{R} \frac{\partial}{\partial R}\right)\Phi\right]_{i,j}
&=&
\begin{cases}
\displaystyle\frac{4\left(\Phi_{2,j} - \Phi_{1,j}\right)}{\delta^2} & (i=1)\\
\displaystyle\frac{\Phi_{i+1,j} - 2\Phi_{i,j} + \Phi_{i-1,j} }{\delta^2}  + \displaystyle\frac{1}{R_i} \frac{ \Phi_{i+1,j} - \Phi_{i-1,j} }{2\delta} & (1 < i < N) \\
\end{cases} 
\label{eq:d2phidrho2}\\
\left[\frac{\partial^2}{\partial z^2}\Phi\right]_{i,j}
&=&
\begin{cases}
\displaystyle\frac{2\left(\Phi_{i,2} - \Phi_{i,1}\right)}{\delta^2} & (j=1)\\
\displaystyle\frac{\Phi_{i,j+1} - 2\Phi_{i,j} + \Phi_{i,j-1} }{\delta^2} & (1 < j < N) \\
\end{cases} 
\label{eq:d2phidz2}
\ee
\end{widetext}
Note that we do not need to define $\nabla^2\Phi$ at $i=N$ and/or $j=N$. 

We start by initialising $\Phi$ as zero everywhere, assigning an initial test profile of $\chi$ that is conveniently chosen as some numerical approximation of the known self-gravitating solution; see, e.g.~\cite{Marsh:2015wka}. Throughout the calculation we enforce
\be\chi_{i,N} = \chi_{N,i} &=& 0. \label{eq:BCchi}\ee

The discretised Eq.~(\ref{eq:poisson}) is then solved iteratively using the  successive over-relaxation (SOR) method (see, e.g., ch.19.5 in~\cite{numericalrecipes}). In each iteration of the program, 
$\Phi_{i,j}\,(i,j\neq 1,N)$ is improved by the SOR method as
\be
\Phi_{i,j}^{\rm new} &=&
\Phi_{i,j}^{\rm old} +\frac{\omega_\Phi \delta^2}{4}\,\times \no\\
&&\biggl(
\frac{\Phi_{i+1,j}^{\rm old} - 2\Phi_{i,j}^{\rm old} + \Phi_{i,j-1}^{\rm new} }{\delta^2}%\no\\
%
%& +&
 +\frac{ \Phi_{i+1,j}^{\rm old} - \Phi_{i-1,j}^{\rm new} }{2\delta R_i} \no\\
&+& \frac{\Phi_{i,j+1}^{\rm old} - 2\Phi_{i,j}^{\rm old} + \Phi_{i,j-1}^{\rm new}}{\delta^2} - \chi_{i,j}^2
\biggr), \label{eq:improvement_phi}
\ee
where $\omega_\Phi$ is an auxiliary parameter\footnote{\label{fn:omphi}To obtain our results in this paper, we have used $\omega_\Phi=1.6$ in all computations. This value was chosen somewhat arbitrarily based on tests of the rate of convergence.} that we set ${\cal O}(1)$. 
For $i=1$ and/or $j=1$, the RHS of Eq.~(\ref{eq:improvement_phi}) should be modified according to Eqs.~(\ref{eq:d2phidrho2}-\ref{eq:d2phidz2}). At $i=N$ and/or $j=N$, $\Phi_{i,j}$ is fixed by the following boundary conditions,
\be
\Phi_{i,N} = \Phi_{N,i} &=& -\frac{\tilde M}{4\pi r_{i,N}}. \label{eq:BCphi} 
\ee
Here, $r_{i,N} = \sqrt{1 + (i-1)^2/(N-1)^2}L$ and the dimensionless\footnote{$\tilde M$ is related to $M$, the physical mass of the soliton, via $M=\left(\frac{M_{pl}^2}{4\pi m}\right)\tilde M$.} $\tilde M$ is calculated as
\be
 \tilde M &=& \frac{\pi\,\delta^3}{4}\,\chi_{1,1}^2 + \left(\sum_{i=2}^{N}2\pi R_i\,\delta^2\,\chi_{i,1}^2\right) + \left(\sum_{j=2}^{N} \frac{\pi\,\delta^3}{2} \chi_{1,j}^2\right) \no \\
 	&+& \sum_{i,j=2}^{N} 4\pi\,R_i\,\delta^2\,\chi_{i,j}^2,
\ee
consistent with Gauss' Law (see Sec.~\ref{app:bh} below).

Next, once $\Phi$ is fixed, the ground state solution of Eq.~(\ref{eq:schrodinger}), $\chi_0$, can be found by considering the following imaginary time evolution (see also App.~\ref{app:theory}):
\begin{align}
\frac{\partial}{\partial\tau} \chi(\tau) = \nabla^2 \chi - 2\left(\Phi + \Phi_b\right)\chi(\tau).
\end{align}
In the large $\tau$ limit, the asymptotic behaviour of $\chi$ is
\begin{align}
\lim_{\tau\to\infty}\chi(\tau) \propto e^{-2\gamma \tau} \chi_0.
\end{align}
Thus, in each iteration, 
$\chi_{i,j} (i,j\neq 1,N)$ is improved as
\be
\tilde\chi_{i,j} &=&
\chi_{i,j}^{\rm old} \no\\
&+&
\frac{\omega_\chi \delta^2}{4} \biggl(
\frac{\chi_{i+1,j}^{\rm old} - 2\chi_{i,j}^{\rm old} + \chi_{i,j-1}^{\rm old} }{\delta^2} \no\\
&&  \qquad\quad + \frac{1}{R_i} \frac{ \chi_{i+1,j}^{\rm old} - \chi_{i-1,j}^{\rm old} }{2\delta} \no\\
&&  \qquad\quad + \frac{\chi_{i,j+1}^{\rm old} - 2\chi_{i,j}^{\rm old} + \chi_{i,j-1}^{\rm old}}{\delta^2} \no\\
&&  \qquad\quad - 2\left(\Phi_{i,j} + \Phi_{b,i,j}\right) \chi_{i,j}^{\rm old} 
\biggr),
\label{eq:improvement_chi}\\
\chi_{i,j}^{\rm new} &=& \frac{\chi_{1,1}^{\rm old}}{\tilde\chi_{1,1}} \tilde\chi_{i,j},
\ee
where $\omega_\chi$ is an auxiliary parameter\footnote{To obtain our results in this paper, we have used $\omega_\chi=0.8$ in all computations. This value was chosen somewhat arbitrarily based on tests of the rate of convergence. We note that setting $\omega_\chi<\omega_\Phi$ appears to be useful (see footnote~\ref{fn:omphi}).} that we set ${\cal O}(1)$. 
For $i=1$ and/or $j=1$, the RHS of Eq.~(\ref{eq:improvement_chi}) should be modified according to the prescription in Eqs.~(\ref{eq:d2phidrho2}-\ref{eq:d2phidz2}). At $i=N$ and/or $j=N$, $\chi_{i,j}$ is fixed by Eq.~(\ref{eq:BCchi}).

We repeatedly update $\Phi$ and $\chi$, using Eqs.~(\ref{eq:improvement_phi}) and~(\ref{eq:improvement_chi}), until convergence is attained.  
The eigenvalue $\gamma$ is calculated as
\be
\gamma &=&
\frac{-1}{2\chi_{i,j} } \biggl(
\frac{\chi_{i+1,j} - 2\chi_{i,j} + \chi_{i,j-1} }{\delta^2}
+ \frac{1}{R_i} \frac{ \chi_{i+1,j} - \chi_{i-1,j} }{2\delta} \no\\
& +& \frac{\chi_{i,j+1} - 2\chi_{i,j} + \chi_{i,j-1}}{\delta^2} - 2\left( \Phi_{i,j} + \Phi_{b,i,j} \right) \chi_{i,j}
\biggr).\no\\&&
\ee
To calculate the total soliton energy, we use Eq.~(\ref{eq:E}) (averaging over adjacent grid sites can be useful in order to reduce numerical error):
\be \label{eq:Ediscrete}
 \tilde{E} &=& \int d^3x\left(\frac{1}{2}\left(\nabla \chi\right)^2+\left(\dfrac{\Phi}{2}+\Phi_b\right)\chi^2\right)\no\\ &=& \sum_{i,j = 1}^{N-1} 2\pi\delta \left(R_{i+1}^2 - R_i^2\right)\no\\ &&\times  \frac{\left[e_{i,j} + e_{i,j+1} + e_{i+1,j} + e_{i+1,j+1}\right]}{4}
\ee
with the integrand 
\be \label{eq:Eint}
 e_{i,j} &=& \frac{-\chi_{i,j}}{2} \biggl(
\frac{\chi_{i+1,j} - 2\chi_{i,j} + \chi_{i,j-1} }{\delta^2}
+ \frac{1}{R_i} \frac{ \chi_{i+1,j} - \chi_{i-1,j} }{2\delta} \no\\
& +& \frac{\chi_{i,j+1} - 2\chi_{i,j} + \chi_{i,j-1}}{\delta^2} - \left( \Phi_{i,j} + 2\Phi_{b,i,j} \right) \chi_{i,j}\biggr).\no\\&&
\ee
For $i=1$ and/or $j=1$, the RHS of Eq.~(\ref{eq:Eint}) should be modified according to the prescription in Eqs.~(\ref{eq:d2phidrho2}-\ref{eq:d2phidz2}). We do not need to include $i=N$ and/or $j=N$, because there the integrand vanishes due to the boundary condition in Eq.~\eqref{eq:BCchi}.

This concludes the description of the numerical scheme.

\subsection{Adding a black hole}\label{app:bh}

Here we explain how a central black hole (BH) can be added to the discretised grid calculation of App.~\ref{app:num}.
To this end we derive a discretised version of Gauss's Law. 

Using Eqs.~(\ref{eq:d2phidrho2}-\ref{eq:d2phidz2}), for $n<N$, we obtain
\be
\sum_{i=1}^n \kappa_i \delta \left[\left( \frac{\partial^2}{\partial R^2} + \frac{1}{ R}\frac{\partial}{\partial R} \right) \Phi\right]_{i,j}
&=& \pi (2n-1) \frac{ \Phi_{n+1,j}- \Phi_{n,j}}{\delta}, \no\\&&\\
\sum_{j=1}^n \lambda_j \delta \left( \frac{\partial^2  \Phi}{\partial z^2} \right)_{i,j} 
&=& 2\frac{ \Phi_{i,n+1}- \Phi_{i,n}}{\delta},
\ee
where $\kappa_i$ and $\lambda_j$ are defined as
\begin{align}
\kappa_i = \begin{cases}
\pi/4 & (i=1) \\
2(i-1)\pi & (i\neq 1)
\end{cases}, \quad
\lambda_j = \begin{cases}
1 & (j=1) \\
2 & (j\neq 1)
\end{cases}.
\end{align}
From these equations we obtain 
\be
&& \sum_{i=1}^{n_i} \sum_{j=1}^{n_j} \kappa_i \lambda_j \delta^3 (\nabla^2  \Phi)_{i,j} \no\\ &=&
  2\sum_{i=1}^{n_i} \kappa_i \delta^2 \times \frac{ \Phi_{i,n_j+1} -  \Phi_{i,n_j}}{\delta} \no\\
&+& 2\pi \sum_{j=1}^{n_j} \lambda_j \frac{2n_i-1}{2} \delta^2 \times \frac{ \Phi_{n_i+1,j} -  \Phi_{n_i,j}}{\delta}, 
\label{eq:gausslaw}
\ee
where $n_i,n_j<N$.
This becomes the usual Gauss's law $\int_V \nabla^2 \Phi = \int_{\partial V} d  S \cdot   \nabla  \Phi$ in the limit $\delta \to 0$.

Consider a black hole with physical mass $M_{\rm BH}$, translated in our conventions to $M_{\rm BH}=(M_{pl}^2/4\pi m)\tilde M_{\rm BH}$. It gives the potential $\Phi_{\rm BH} = -\tilde M_{\rm BH}/(4\pi x)$.
The Poisson equation is $\nabla^2  \Phi = \rho$; thus, the discretised $\rho$ configuration that leads to $\Phi_{\rm BH}$ and that is consistent with Gauss's Law, Eq.~(\ref{eq:gausslaw}), is
\be
\rho_{i,j}
&=&
\begin{cases}
\frac{4 \tilde M_{\rm BH}}{\pi \delta^3} & (i = j = 1 ) \\
0                & (i\neq 1~{\rm or}~j\neq 1 )
\end{cases}.\label{eq:BHrho}
\ee

Finally, to interface to the code described in App.~\ref{app:num}, it is convenient to utilise the gravitational potential induced by the BH which is given on the axisymmetric grid as follows,
\be
\Phi_{{\rm BH},i,j}
&=&
\begin{cases}
-\frac{11 \tilde M_{\rm BH}}{12\pi \delta} & (i = j = 1 ) \\
-\frac{\tilde M_{\rm BH}}{4\pi r_{i,j}}                & (i\neq 1~{\rm or}~j\neq 1 )
\end{cases}.
\ee
The tricky point here is $\Phi_{{\rm BH},1,1}$: this is determined by solving the discretised Poisson equation at the origin.

\section{Is it the ground state?}\label{app:theory}
In this section we present evidence that these solutions obtained by our algorithm are indeed ground state solutions. At the same time, we also highlight the difficulty to obtain a rigorous proof. Finally, we comment about stability to perturbations.

Let us recap a few details of the imaginary time evolution of the Schr\"odinger equation \cite{Grimm1969,Goldberg1967a,Goldberg1967b}. The equation reads
\be
i\partial_t \Psi(\mathbf{r},t) = H\Psi(\mathbf{r},t) .
\ee
The time-independent Hamiltonian $ H $ is hypothesized to have eigenfunctions $ \phi_n(\mathbf{r}) $ with eigenvalues $ \epsilon_n $, including a ground-state with finite $ \epsilon_0<0 $. 

Consider an initial condition
\be
\Psi(\mathbf{r}, 0) = \sum_n a_n\psi_n(\mathbf{r}),
\ee
which can be propagated in time as
\be
\Psi(\mathbf{r},t) = \sum_n e^{-i\epsilon_n t}a_n\psi_n(\mathbf{r}).
\ee
We can define $ \tau = it $ and rewrite the Schr\"odinger equation as
\be
\partial_\tau \widetilde{\Psi}(\mathbf{r},\tau) = -H\widetilde{\Psi}(\mathbf{r},\tau),
\ee
with the initial condition
\be
\widetilde{\Psi}(\mathbf{r}, 0) = \sum_n a_n\psi_n(\mathbf{r})
\ee
and a general solution
\be
\widetilde{\Psi}(\mathbf{r},\tau) = \sum_n e^{-\epsilon_n \tau}a_n\psi_n(\mathbf{r}).
\ee
In the limit $ \tau\to \infty $, we have:
\be\label{eq:gs}
\lim_{\tau\to\infty} \widetilde{\Psi}(\mathbf{r},\tau)  = e^{-\epsilon_0 \tau} a_0 \psi_0(\mathbf{r}) ,
\ee
thus providing the sought-after ground-state, $ \psi_0 $. 

The difficulty in this formalism, which becomes apparent in the regime where the self-gravitation is dynamically relevant, is that the Hamiltonian is not constant between iterations but rather changes as we iterate on the wave function and the Newtonian potential induced by it. Thus, while the solutions found by our solver are (within the numerical accuracy) indeed solutions of the EOM, we have no rigorous proof that these are in fact the ground state solutions. Having made this cautionary remark, we now present some evidence that our solution is indeed the ground state, at least when it comes to ULDM in the background baryonic potential of realistic galaxies. 

The first thing to note is that in the limit that self-gravity is negligible compared to the external potential, the problem becomes linear, the Hamiltonian is constant and the derivation leading to Eq.~(\ref{eq:gs}) is applicable without particular complications. Then, the formalism leading to Eq.~(\ref{eq:gs}) suggests that our solution does indeed isolate the true ground state, as long as the initial test function has some non-vanishing overlap with this ground state. In specific examples we can compare the numerical results to analytic solutions. The case of a strong baryonic potential concentrated near the origin is a good example:  in this case the exact solution converges to the Coulomb wave function $\chi(r)\propto e^{-Ar}$. 

On the other hand, in the opposite limit where the external potential is negligible and self-gravity dominates, we find that our algorithm converges to the known self-gravitating ground state solution. 

Many examples in the paper (e.g., Figs.~\ref{fig:MlamMW},\ref{fig:MKMW},\ref{fig:MvsLamugc1281},\ref{fig:EoverMugc1281},\ref{fig:MvsLamf571},\ref{fig:EoverMf571}) explicitly examine the behaviour of the solution while going smoothly between the two limits of negligible external potential and all the way to where the external potential dominates the solution. The two limits are smoothly connected by a continuous deformation. This lends support to the notion, that also in the intermediate regime our solver is finding the true ground state solution. 

We have also made sure that the solutions are not sensitive to the details of the test function used as initial condition. For a given external potential $\Phi_b$, we checked a variety of initial conditions of the field $\chi_{i,j}$, including gaussian forms with different slopes as well as randomised independent realisations of the field on different grid points (always keeping $ \chi_{i=N,j}=\chi_{i,j=N}=0 $ as prescribed in Eq.~(\ref{eq:BCchi})). For some of these initial conditions the solver converges on a solution, while for others it does not. Importantly, whenever the solver does converge, the different initial conditions all lead to the same solution. Note that the solver sometimes does not converge when the initial conditions do not fall steeply enough as a function of distance away from the origin. In addition, convergence also shows some dependence on the function $\Phi_b$ used in the test. A simple choice which works well for all of the problems we experimented with, was to use the spherical self-gravitating solution as the initial test function.

We now make a short comment about the linear stability of our solutions. It is useful to first recall the stability argument for the self-gravitating soliton: in the Newtonian limit, the mass and the energy of the field are conserved separately; since the soliton is the field configuration that minimises the energy at fixed value of the mass, it is guaranteed on general grounds to be dynamically stable~\cite{Chavanis:2011zi}\footnote{Relativistic corrections do cause soliton decay~\cite{Mukaida:2016hwd,Eby:2018ufi}, but the decay time is long and of no phenomenological relevance in the range of ULDM and soliton masses considered in this work.}. 
Once we ``turn on" an external potential (spherically symmetric or not), however, linear stability could become a concern. 

Our solver finds solutions of the EOM while holding the value of the field fixed at the origin (via the $\chi(0)=\lambda^2$ prescription). Let us assume, based on the arguments given in the previous paragraphs, that the solution we find is indeed the lowest energy solution compatible with the boundary conditions. One can show that both in the self-gravitating case, and in case that the external potential fully dominates the dynamics, the value of $\chi(0)$ is in one-to-one correspondence to the mass of the ground state. Therefore we expect that our solutions remain minimisers of the ULDM energy at fixed mass also in the presence of the external potential. This settles the stability question for the limit where the external potential strongly dominates: as long as the external potential is static and does not rearrange itself dynamically following a change in the ULDM system, there is no energy exchange between the ULDM and the external system and the soliton stability is guaranteed.

The intermediate limit, where the external potential is comparable to that coming from the ULDM, is much more complicated. Addressing the question of stability in this case requires a joint analysis of the baryonic system and the ULDM. This analysis is beyond the scope for the current paper. Indeed, our goal in this work is not to solve the (in general, very difficult) dynamical problem of finding stable gravitating solutions of the joint baryonic and ULDM system. Instead, our starting point is to assume that the baryonic part of the system is already known via observational constraints like stellar luminosity and gas line emission (as was the case for the LSB galaxies we analysed), and then derive the minimum energy soliton consistent with this known external background.

\section{Gravitational potential of an axisymmetric mass distribution}\label{app:disk}
The solution of the Poisson equation in axisymmetry can be found directly using the method of Fourier-Bessel transform. Following \cite{Casertano1983a}, the gravitational potential is given by
\be
\phi(R,z) &=& -2\pi G\int\limits_{-\infty}^{\infty}d\zeta\int\limits_{0}^{\infty}du \rho(u,\zeta)K(R,u,|z-\zeta|),\no\\&&
\ee
where the kernel $K$ is given by
\be
K(R,u,z) & =& u\int\limits_0^\infty dk J_0(kR)J_0(ku)e^{-kz} \\ 
& =& \dfrac{\sqrt{u}}{\pi\sqrt{R}}\text{Re}\left[Q_{-\frac{1}{2}}\left(\dfrac{R^2+u^2+z^2}{2Ru}\right)\right]\no
\ee
with $Q_{-\frac{1}{2}}$ the Legendre function of the second kind of order $-\frac{1}{2}$. 
See also 6.612 (3) and 8.834 (1) in Ref.~\cite{GradshteynRyzhik}.

\section{Modelling stellar disks}\label{app:mnfits}
In order to simplify the analysis we take advantage of the Miyamoto-Nagai (MN)~\cite{Miyamoto1975a} disk parametrisation, described by the density profile
\be
\rho_{MN}(R,z)   
& =&\frac{M_{MN}b^2}{4\pi}\,\times\no\\
&&\frac{aR^2+\left(a+3\sqrt{z^2+b^2}\right)\left(a+\sqrt{z^2+b^2}\right)^2}{\left(R^2+(a+\sqrt{z^2+b^2})^2\right)^{5/2}\left(z^2+b^2\right)^{3/2}}\no\\&&
\ee
for which the gravitational potential is known analytically:
\be
\phi_{MN}(R,z)&=&-\dfrac{GM_{MN}}{\sqrt{R^2+(a+\sqrt{z^2+b^2})^2}}.
\ee
The parameters $a,\,b,$ and $M_{MN}$ define the disk scale-radius, thickness and mass. 
A sum of three MN profiles provides a reasonable approximation to the   exponential disks of typical galaxies~\cite{Smith2015}. 

In the SPARC database the surface brightness $\Sigma_L(R)$ of the disk component is reported. Assuming an exponential approximation for the vertical direction, the stellar mass density is given by
\be
\rho(R,z) &=& \Upsilon_d \Sigma_L(R)\dfrac{e^{-\frac{|z|}{z_d}}}{2z_d}
\ee
where $z_d$ is specified in the database for each galaxy. 
One can then fit the MN density on the disk plane,
\be
\rho_{MN}(R,0)&=&\dfrac{M_{MN}\left[aR^2+(a+3b)(a+b)^2\right]}{4\pi b\left[R^2+(a+b)^2\right]^{5/2}}, 
\ee
(or a sum of such functions) to
\be
\rho(R,0)&=&\dfrac{\Upsilon_d\Sigma_L(R)}{2z_d},
\ee
fixing $b=z_d$.  

\begin{figure}[t!]
	\centering
	\includegraphics[width=0.495\textwidth]{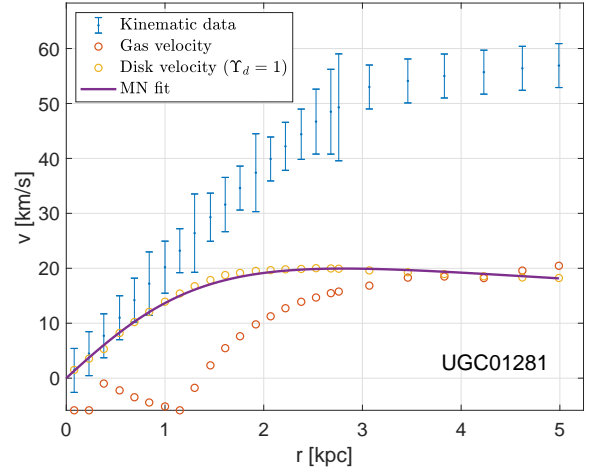}
	\caption{MN fit for the disk of UGC01281. The disk fit is shown by the solid purple line, while the orange circles show the velocity attributed to the disk in the SPARC database for $\Upsilon_d=1$.
	}\label{fig:MNrcs1}
\end{figure}

In Fig.~\ref{fig:MNrcs1} we show the rotation curve decomposition for UGC01281, superimposed with a MN fit for the disk obtained with the above prescription. 

\section{Modelling the gas distribution in UGC01281}\label{app:ugc1281gas}

The SPARC database~\cite{Lelli:2016zqa} does not contain sufficient information to allow a direct reconstruction of the gas mass distribution\footnote{We thank Stacy McGaugh for clarifications on this point.}. We have therefore done an independent analysis of the gas component for the example of UGC01281, using the HI surface brightness profiles reported in Ref.~\cite{Kamphuis:2011qg}. 

Our analysis is less sophisticated than that in~\cite{Kamphuis:2011qg}, but captures the key features of the gas profile with sufficient accuracy. We model the gas density profile as a collection of $K$ co-planar rings, with the mass density of each ring taking to be constant on the plane ($z=0$) and decaying vertically with a Gaussian profile:
\be\rho_{\rm gas}\left(R,z\right)&=&\sum_{k=1}^K\theta\left(R-R_k\right)\theta\left(R_k+\Delta_k-R\right)\rho_k\,e^{-\frac{z^2}{d_z^2}},\no\\&&\ee
where $\theta(x)$ is the Heaviside function. The gravitational potential due to this mass distribution is computed by the procedure given in App.~\ref{app:disk}.

The surface brightness profile from this gas distribution is easily computed. Matching the model to the vertical profile reported in~\cite{Kamphuis:2011qg}, we find a good fit for $d_z=0.65$~kpc. Considering the radial profile and matching (approximately, by eye) to the average profile shown in Fig.~2 of Ref.~\cite{Kamphuis:2011qg} (which averages the HI column density over a slab in the vertical direction), with find that a model of $K=50$ rings of equal width $\Delta_k=0.2$~kpc, located with inner radii starting at $R_1=0$~kpc up to $R_{50}=10$~kpc, reproduces the brightness profile radial shape for the density assignment $\rho_k=\tilde\rho\left(0.5+R_k\right)^{1.2}\exp\big({-\left(\frac{R_k}{1.5}\right)^{1.4}}\big)$, where $R_k$ are noted in kpc and $\tilde\rho$ is an over-all normalisation factor. We set $\tilde\rho=3.9\times10^6$~M$_\odot$/kpc$^3$, so that the total gas mass (including a factor of 1.3 to account for He) is fixed to $M_{\rm gas}=3.2\times10^8$~M$_\odot$, inferred in Ref.~\cite{deBlok:2002vgq} from the total HI luminosity. 

\begin{figure}[htbp!]
\centering
\includegraphics[width=0.495\textwidth]{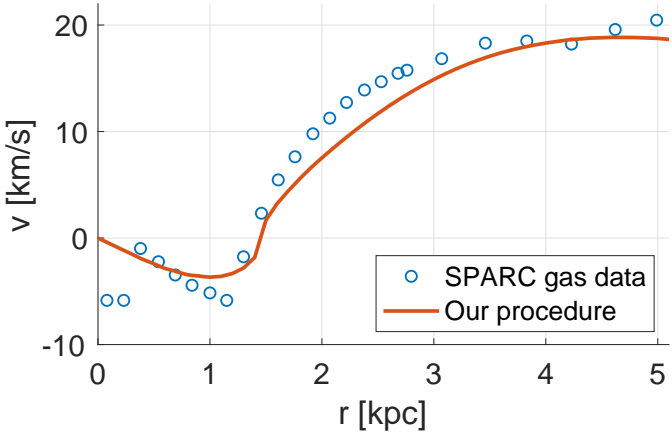}
\caption{Modelling the gas contribution to the rotation curve of UGC01281.}\label{fig:UGC01281Gas}
\end{figure}

The gas-induced rotation curve we find with this procedure is shown by the line in Fig.~\ref{fig:UGC01281Gas}, compared to the velocity contribution attributed to the gas in the SPARC database (circles). The comparison is good enough for our purpose in the current work: as we show in the body of the work, the total baryonic effect (stars and gas combined) on the predicted soliton and on the large-scale halo of UGC01281 is small. 

\begin{figure}[htbp!]
\centering
\includegraphics[width=0.5\textwidth]{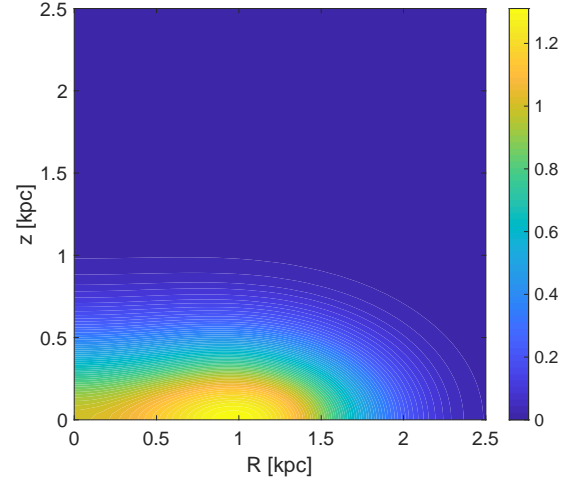}
\caption{Density profile of a toroidal soliton solution.}\label{fig:TorusSoliton}
\end{figure}

We conclude this technical discussion with an amusing comment. The toroidal gas profile of UGC01281 prompted us to look for toroidal soliton solutions, that could co-exist in the background potential of such a baryonic mass distribution. Indeed, varying the gas mass and the soliton mass, we can find toroidal solitons; we show an example in Fig.~\ref{fig:TorusSoliton}. The parameters chosen to achieve this toroidal solution were: $m=10^{-22}$~eV, with $\lambda=10^{-5}$ and a gas mass 50 times larger than the observed one in UGC01281. These parameters do not represent an actual galaxy from SPARC: we merely bring it as an observation about deformed solitons and as demonstration of the versatility of the numerical code.

\section{Results with ULDM particle mass of $m=10^{-21}$~eV}\label{app:1e-21}

Here we present a repetition of Figs.~\ref{fig:MvsLamugc1281}-\ref{fig:EoverMf571} from Sec.~\ref{s:sparc}, done for ULDM particle mass $m=10^{-21}$~eV. 
\begin{figure}[htbp!]
	\centering
	\includegraphics[width=0.495\textwidth]{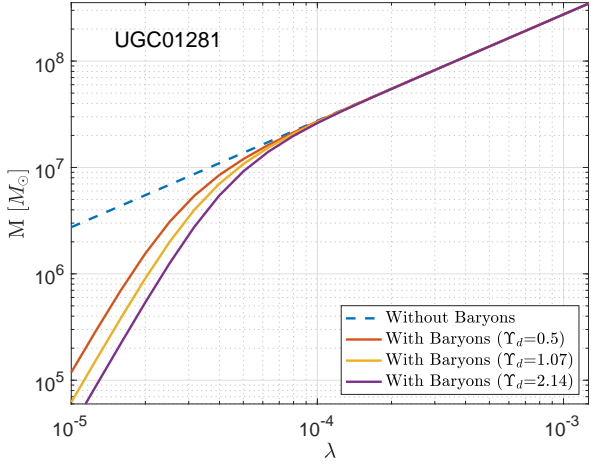}
	\includegraphics[width=0.495\textwidth]{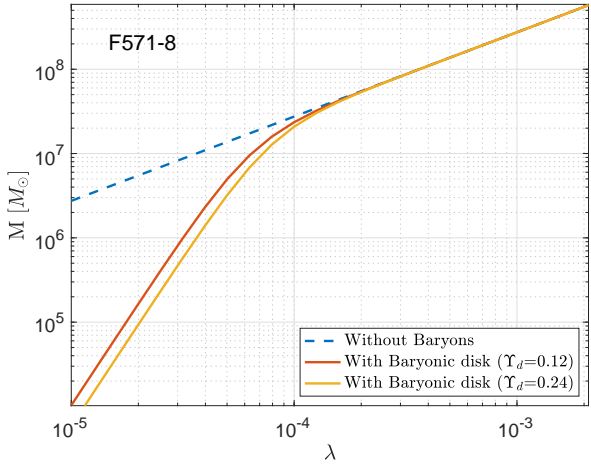}
	\caption{Soliton M-$\lambda$ relation in baryonic-induced background gravitational potential of SPARC galaxies. Top panel: UGC01281. Bottom panel: 571-8. The ULDM particle mass is $m=10^{-21}$~eV.}\label{fig:MvsLamugc1281_21}
\end{figure}
\begin{figure}[htbp!]
\centering
\includegraphics[width=0.495\textwidth]{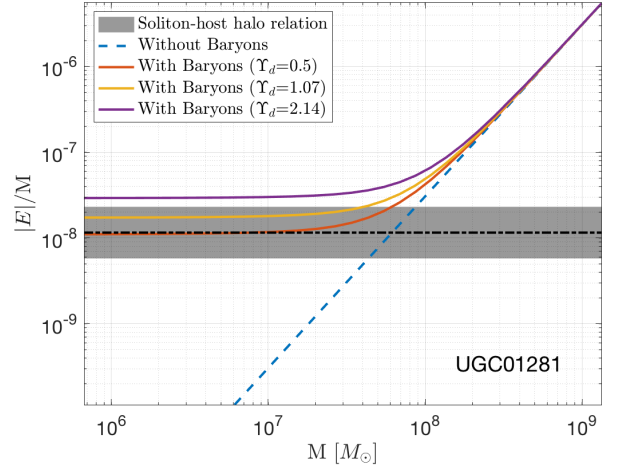}
\includegraphics[width=0.495\textwidth]{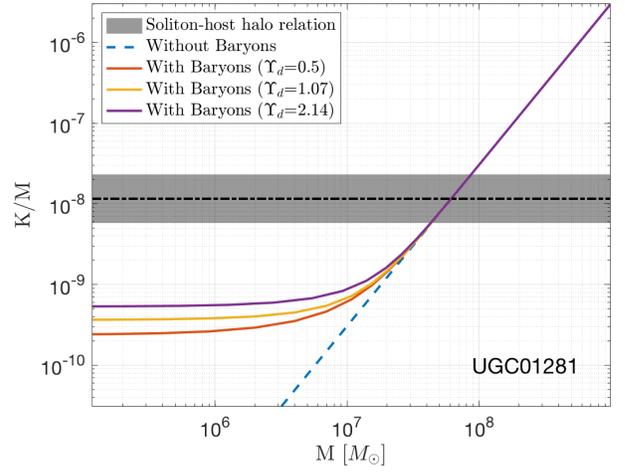}
\caption{Specific energy $|E|/M$ (top) and specific kinetic energy $K/M$ (bottom) for a soliton in UGC01281. The soliton host-halo relation found in DM-only numerical simulations is shown by the black dashed line with a shaded band denoting a factor of two estimated spread. The ULDM particle mass is $m=10^{-21}$~eV.}\label{fig:EoverMugc1281_21}
\end{figure}
\begin{figure}[htbp!]
\centering
\includegraphics[width=0.495\textwidth]{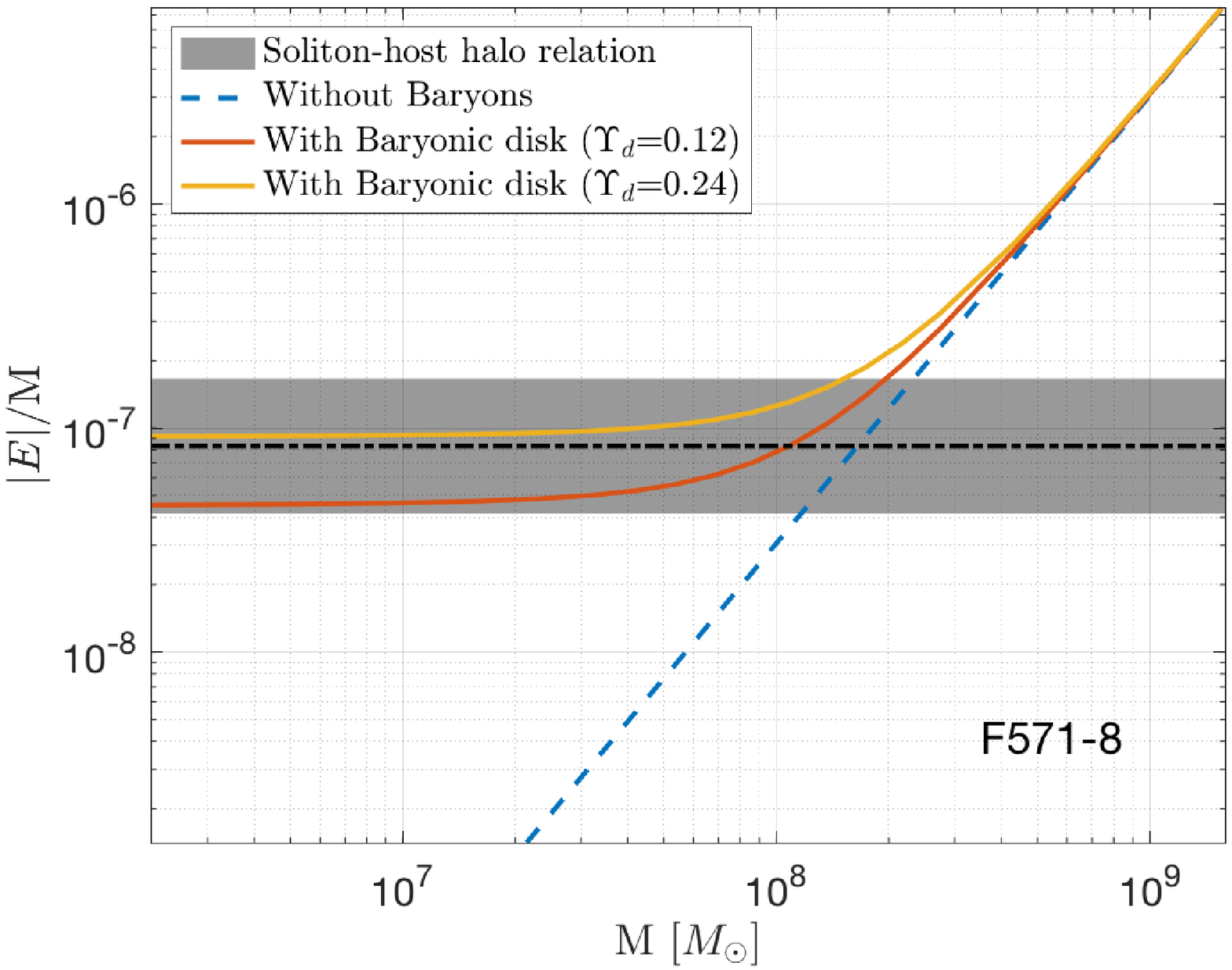}
\includegraphics[width=0.495\textwidth]{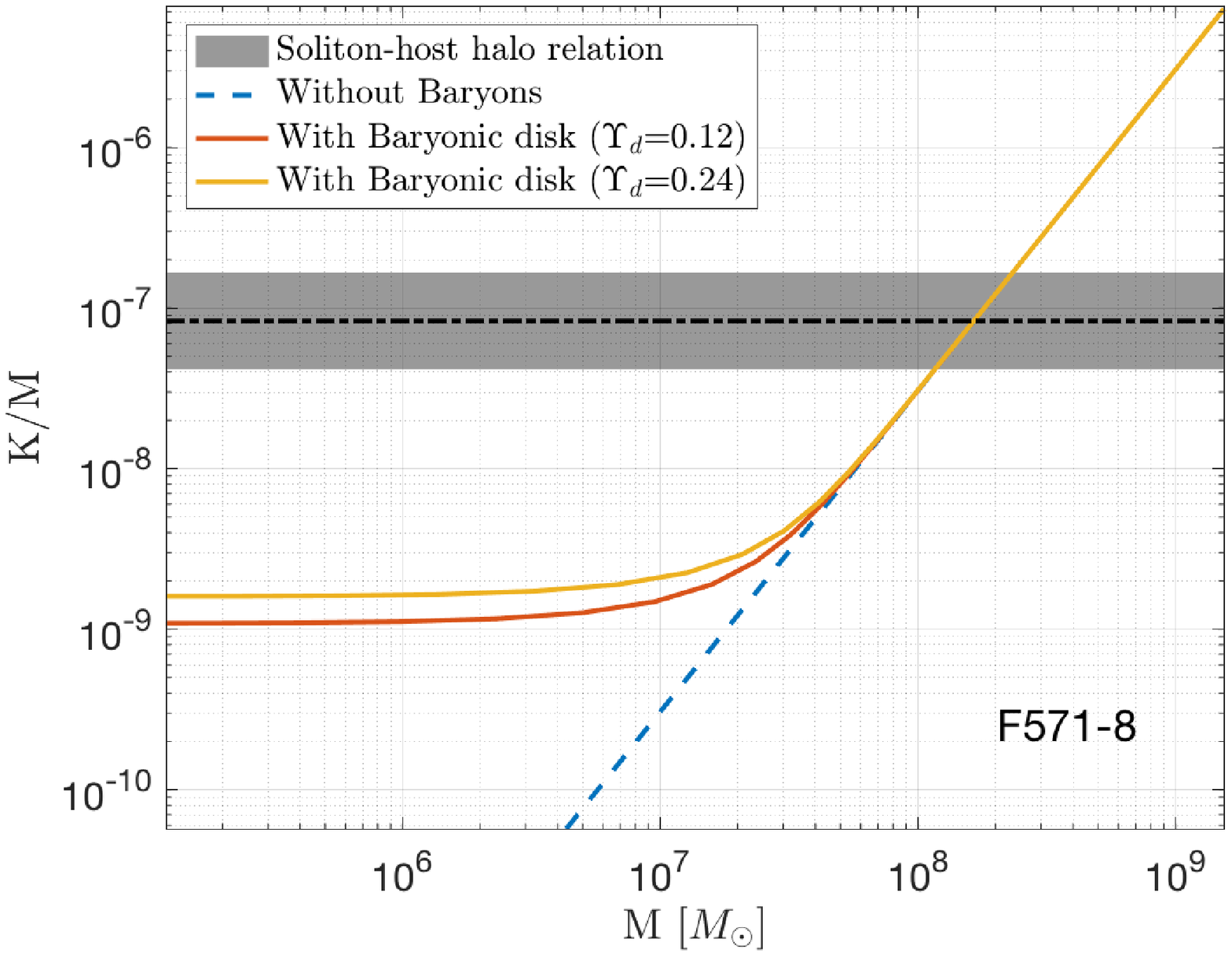}
\caption{Same as Fig.~\ref{fig:EoverMugc1281_21}, but done for F571-8.}\label{fig:EoverMf571_21}
\end{figure}

\end{appendix}
%\vspace{6 pt}
%\newpage
\clearpage

\bibliography{ref}
\bibliographystyle{utphys}

\end{document}